# DOA Estimation in Antenna Arrays: Super-Resolution Methods in a Common Space

Sergey S. Kuzin

*Abstract*— **This paper introduces a unified framework for angular super-resolution in antenna arrays. By integrating linear prediction beyond the physical aperture, entropy variations, and harmonic spatial spectra averaging, we recast classical methods—including Capon, Pisarenko, Maximum Entropy, and MUSIC—into a single structure. We extend this framework to accommodate ESPRIT-type methods by treating the number of simultaneously predicted subarray channels as an additional dimension. Furthermore, we adapt this approach to non-Gaussian signals, establishing a parallel framework for Virtual-ESPRIT-type methods. As a practical application, we propose universal, J-orthogonalization-based algorithms that estimate spatial super-resolution spectra within a unified computational architecture. This architecture produces spatial spectrum estimates for multiple super-resolution methods and any intermediate variations.**

## I. Introduction

Direction of Arrival (DOA) estimation is a foundational technique in array signal processing, with critical applications in radar, sonar, and modern wireless communications [1]. To achieve higher angular resolution beyond traditional limits, researchers have introduced various super-resolution (sub-Rayleigh) algorithms. Prominent examples include Maximum Entropy (ME) [2], Capon [3], Pisarenko [4], MUSIC [5], ESPRIT [6], and VESPA [7]. These methods usually fall into two views: linear prediction or signal-noise subspaces. However, strong links exist within and between these groups. This paper presents a common view of angular super-resolution methods as linear prediction of signals beyond an antenna aperture. This framework shows that shifting between methods simply involves changing one or more of these three parameters. It also allows us to derive new estimates that lie anywhere between existing methods.

This approach enables universal algorithms to produce spatial spectrum estimates across multiple methods. We demonstrate this by using the J-orthogonalization algorithm to obtain estimates of the Pisarenko, maximum entropy, Capon and MUSIC spectra, as well as any intermediate estimates positioned between these methods.

The paper is organized as follows. Section 2 examines angular super-resolution methods using linear signal prediction beyond an antenna aperture, ranging from maximum entropy (ME) to minimum entropy (Pisarenko).

The author is with National Research University Higher School of Economics (HSE), Moscow, Russia (e-mail: ss.kuzin@hse.ru).

Section 3 examines multichannel linear prediction using subarrays with varying entropy; within this framework, the minimum and maximum entropy approaches correspond to the original and approximate ESPRIT methods, respectively.

In Section 4 we demonstrate that VESPA-type super-resolution methods—an extension of EXPRIT for non-Gaussian signals—can be considered as liner prediction of signals beyond both real and virtual antenna arrays.

Section 5 shows that the Capone spatial spectrum results from the harmonic averaging of maximum entropy estimates. Similarly, the MUSIC spectrum results from the harmonic averaging of Pisarenko estimates

Section 6 introduces a three-dimensional common space for angular super-resolution methods based on a three-dimensional space: the entropy of the correlation function beyond the antenna array aperture, the presence of harmonic averaging, and the use of multichannel signal prediction by subarrays. This space accommodates all super-resolution methods discussed in Sections 3 to 5, along with any intermediate estimators. Two currently unoccupied boundary points in this space correspond to the multichannel variants of Capon and MUSIC. To illustrate the utility of this framework, we synthesize an estimator for one of these missing points—multichannel MUSIC. Additionally, we introduce a corresponding common space for non-Gaussian signal super-resolution.

In Section 7, we propose universal algorithms for estimating spatial super-resolution spectra via J-orthogonalization. Grounded in the relationships between methods within a common space, this unified framework delivers simultaneous spectral estimates for various super-resolution techniques and any intermediate positions between them.

## II. Linear Prediction

### A. Maximum Entropy Linear Prediction

Linear prediction is used to extrapolate signals beyond the physical antenna aperture [8, 9]. No specific behavior may be assumed for the correlation function outside the observation window, other than it exists and remains non-zero. In this case, the function can be extended to maximize randomness while ensuring the resulting function aligns with the measurements within the observation window. This approach corresponds to the Maximum Entropy method [2].

Consider a uniform linear array (ULA) with $N$ antenna elements that are equally spaced with a distance is equal to a half wavelength. Assume that there are $d$ far field narrowband signal sources located at $\theta_1, \theta_2, \ldots, \theta_d$. The received signals and noise are uncorrelated. Under these conditions, the output signal vector of the ULA with the number of snapshots t is as follows:

$$\mathbf{X}(t) = [x_1(t), x_2(t), \dots, x_N(t)]^T = \sum_{k=1}^{d} \boldsymbol{a}(\theta_k) s_k(t) = \mathbf{A}\mathbf{s}(t) + \mathbf{n}(t), \quad (1)$$

where $x_i$ is the output of the $i$'th array element; $\theta_k$ is the DOA of the $k$-th incident signal; $\mathbf{s}(t)$ is the signal vector; $\mathbf{n}(t)$ is the array additive noise vector; and $\mathbf{A}$ is the Vandermonde matrix, $i$'th column of which is the steering vector of the $k$'th signal:

$$\boldsymbol{a}(\theta_k) = \left[1, \exp\left\{j2\pi \frac{\Delta}{\lambda} \sin(\theta_k), \dots, j2\pi(N-1) \frac{\Delta}{\lambda} \sin(\theta_k)\right\}\right]^T, \quad (2)$$

where $\lambda$ is the wavelength of the incident signals, $\Delta$ is the element spacing and is no more than half the wavelength, and the covariance matrix is as follows

$$\mathbf{R} = E\{\mathbf{X}(t)\mathbf{X}(t)^{\boldsymbol{H}}\} = \mathbf{A}\mathbf{S}\mathbf{A}^{\boldsymbol{H}} + \sigma^2 \mathbf{I},$$

where $\mathbf{S} = E\{\mathbf{s}(t)\mathbf{s}(t)^H\}$ is the signal covariance matrix, $\sigma^2$ is an unknown Gaussian noise variance.

The entropy of a stationary random process is given by the expression [10]

$$H = \int_{-\pi}^{\pi} \ln[P(\omega)] d\omega, \quad (3)$$

where $\omega = 2\pi \frac{\Delta}{\lambda} \sin\theta$ is the spatial frequency, and $P(\omega)$ is the spatial spectrum,

$$P(\omega) = \sum_{n=-\infty}^{n=\infty} r_n e^{-j\omega n}, \quad (4)$$

where $r_n$ is a spatial correlation function.

Entropy can be parameterized directly through the determinant of the spatial correlation matrix [10],

$$H = \lim_{N\to\infty} \frac{1}{2} ln[\det\mathbf{R}_N], \quad (5)$$

where $\mathbf{R}_\mathrm{N}$ is a covariance matrix of $N$-element antenna array.

The maximum entropy estimate of the spatial spectrum will be

$$P_{ME}(\theta) = \frac{1}{\mathbf{a}(\theta)^H \mathbf{f}\mathbf{f}^H \mathbf{a}(\theta)} = \frac{1}{|\mathbf{a}(\theta)^H \mathbf{f}|^2}, \quad (6)$$

where $\mathbf{a}(\theta) = [1, e^{j\omega}, e^{j2\omega}, \dots, e^{j(N-1)\omega}]^T$ is the steering vector, $\mathbf{f}$ is the first column of $\mathbf{R}^{-1}$.

DOA estimates correspond to the peak locations of the spatial spectrum (6). The infinite correlation function in (5) with maximum entropy is uniquely determined by the vector f. Because this is an autoregressive process of order $N$, the elements of vector $\mathbf{f}$ serve as the corresponding autoregressive coefficients. Each subsequent value of the correlation function outside the observed window can be determined by the condition that the segment of the correlation function is orthogonal to the vector $\mathbf{f}$. Consequently, estimating the spectrum using the autoregressive coefficients in (6) is equivalent to extrapolating the correlation function of the received signals beyond the antenna aperture using maximum entropy. In other words, maximum entropy spatial correlation function should be as close as possible to spatially white noise while matching the known points within the aperture.

*B. Maximum to Minimum Entropy Linear Prediction*

Maximum entropy approach assumes minimum knowledge about the behavior of the correlation function beyond the antenna array. However, this does not always fit radar surveillance, where some prior information about the correlation function beyond the observed segment is available. For far-field sources, wavefronts at the array can be treated as plane waves and extrapolated beyond the aperture. This approach constitutes a linear prediction framework; however, it optimizes for minimum entropy rather than maximum entropy.

The entropy of the spatial correlation function for Gaussian signals, as determined by expression (5), is uniquely related to the determinant of its covariance matrix. Consequently, the problem of minimizing this entropy is equivalent to minimizing the determinant of the covariance matrix for a stationary process extended to infinity beyond the observed region. In this case, due to the equidistance and stationarity of the spatial and temporal sample, the covariance matrix of signals is Toeplitz.

Extending the spatial covariance function$[r_0, r_1, \dots, r_{N-1}]$ by one step corresponds to finding one new element $r_N$ of the Hermitian Toeplitz covariance matrix where we denote it as $\zeta$:

$$\mathbf{R}_N = \begin{bmatrix} r_0 & r_1^* & \dots & r_{N-1}^* & \zeta^* \\ r_1 & r_0 & \dots & r_{N-2}^* & r_{N-1}^* \\ \dots & \dots & \dots & \dots & \dots \\ r_{N-1} & r_{N-2} & \dots & r_0 & r_1^* \\ \zeta & r_{N-1} & \dots & r_1 & r_0 \end{bmatrix}, \quad (7)$$

ensuring the covariance matrix is semi-definite, $\det(\mathbf{R}_N) \geq 0$.

By applying Sylvester's determinant identity [11], we can express the determinant of the Toeplitz covariance matrix $\mathbf{R}_N$ as a function of the predicted value:

$$\det(\mathbf{R}_N) = D_N(\zeta) = \frac{D_{N-1}^2 - |D_{N-1,(1)}(\zeta)|^2}{D_{N-2}}, \quad (8)$$

where

$$D_{N-1} = \det(\mathbf{R}_{N-1}),$$

$$D_{N-1,(1)}(\zeta) = \det \begin{bmatrix} r_1 & r_0 & \dots & r_{N-2}^* \\ \dots & \dots & \dots & \dots \\ r_{N-1} & r_{N-2} & \dots & r_0 \\ \zeta & r_{N-1} & \dots & r_1 \end{bmatrix}.$$

As $\det(\mathbf{R}_N)$ is a quadratic function of the predicted element $\zeta$, there are two values of $\zeta$, where the matrix determinant is equal to zero. These values define the boundaries for the solutions, ranging from minimum to maximum entropy as shown in Fig. 1.

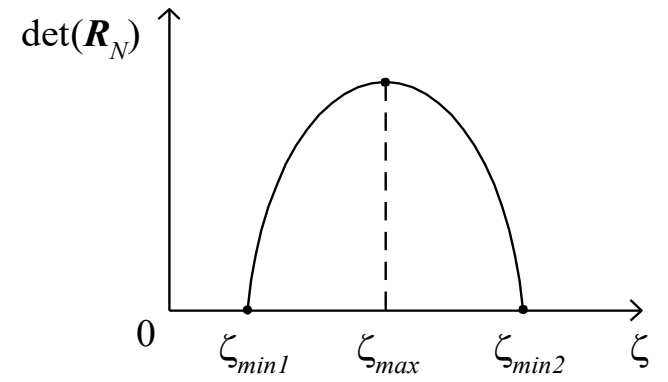

Fig. 1. Minimum and maximum entropy solutions.

Since $D_{N-1}$ and $D_{N-2}$ are independent of $\zeta$, maximizing the determinant of the covariance matrix simplifies to minimizing the minor $D_{N-1,(1)}(\zeta)$. By applying the Sylvester identity to the determinant $D_{N-1,(1)}(\zeta)$, we transform (8) to

$$D_N(\zeta) = \frac{D_{N-1}^2 - |\zeta D_{N-2} + D_{N-1,(1)}(0)|^2}{D_{N-2}}. \quad (9)$$

This means, applying the maximum entropy condition yields the following equation

$$\zeta D_{N-2} + D_{N-1,(1)}(0) = 0, \quad (10)$$

whereas the minimum entropy condition gives the equation

$$D_{N-1}^2 - |\zeta D_{N-2} + D_{N-1,(1)}(0)|^2 = 0. \quad (11)$$

Assuming $D_{N-2} > 0$ and $D_N(\zeta) \geq 0$, we find a single solution of the maximum entropy and two solutions of the minimum entropy ($D_N(\zeta) = 0$):

$$\zeta_{\max} = \frac{D_{N-1,(1)}(0)}{D_{N-2}}, \quad (12)$$

$$\zeta_{\min 1} = \frac{D_{N-1,(1)}(0) - D_{N-1}}{D_{N-2}}, \quad (13)$$

$$\zeta_{\min 2} = \frac{D_{N-1,(1)}(0) + D_{N-1}}{D_{N-2}}. \quad (14)$$

Minimum entropy roots may also be represented as

$$\zeta_{\min 1,2} = \zeta_{\max} \pm \frac{D_{N-1}}{D_{N-2}}. \quad (15)$$

Expressions (12) and (15) determine the next predicted coefficient of the correlation function according to the criterion of maximum and minimum entropy, respectively. Because the extrapolated correlation function depends linearly on the first N known coefficients, the coefficients of a linear filter that generates the entire function can be directly determined. These filter coefficients define a spatial spectrum identical to that of the underlying infinite correlation function. Consequently, the infinite correlation function can be parameterized directly as a vector of linear prediction coefficients.

To derive the vector of linear prediction coefficients, we assume the maximum and minimum entropy correlation coefficients are given by (12) and (13), respectively. Accordingly, we partition the covariance matrix $\mathbf{R}_N$ into the following block form:

$$\mathbf{R}_N = \begin{bmatrix} r_0 & \mathbf{r}_N^H(\zeta) \\ \mathbf{r}_\mathbf{N}(\zeta) & \mathbf{R}_{N-1} \end{bmatrix}. \quad (16)$$

Regardless of the variant of the criterion for extending covariance matrix, maximum or minimum entropy, the vector of coefficients of the linear prediction filter f is determined by the equation

$$\begin{bmatrix} r_0 & \mathbf{r}_N^H(\zeta) \\ \mathbf{r}_\mathbf{N}(\zeta) & \mathbf{R}_{N-1} \end{bmatrix} \begin{bmatrix} 1 \\ \mathbf{f} \end{bmatrix} = \begin{bmatrix} \rho_N \\ \mathbf{0} \end{bmatrix}, \quad (17)$$

where $\rho_N$ is a prediction error variance, which, for the minimum entropy solution, is assumed to be zero.

The columns of the covariance matrix $\mathbf{R}_{N-1}$ in (17) are linearly independent and span the complete $N$-dimensional space containing the vector $\mathbf{r}_N(\zeta)$. Therefore, regardless of the value of $\zeta$, the vector $\mathbf{r}_N(\zeta)$ can be expressed through the columns of the matrix$\mathbf{R}_{N-1}$,

$$\begin{bmatrix} \mathbf{r}_\mathbf{N}(\zeta) & \mathbf{R}_{N-1} \end{bmatrix} \begin{bmatrix} 1 \\ \mathbf{f} \end{bmatrix} = \mathbf{0}, \quad (18)$$

hence

$$\mathbf{f} = -\mathbf{R}_{N-1}^{-1} \mathbf{r}_\mathbf{N}(\zeta). \quad (19)$$

Substituting $\zeta \in [\zeta_{\min 1}, \zeta_{\max}]$ (or $\zeta \in [\zeta_{\max}, \zeta_{\min 2}]$) defines linear prediction coefficient vectors spanning the maximum, minimum, and all intermediate entropy solutions.

For the *maximum entropy* solution $D_{N-1,(1)}(\zeta_{max}) = 0$. It means there is a vector **f** such that

$$\begin{bmatrix} \mathbf{r}_{N-1} & \mathbf{R}_{N-2} \\ \zeta_{\max} & \mathbf{r}_{N-1}^H \bar{\mathbf{J}} \end{bmatrix} \begin{bmatrix} 1 \\ \mathbf{f} \end{bmatrix} = \begin{bmatrix} \mathbf{0} \\ 0 \end{bmatrix} \quad (20)$$

where $\bar{\mathbf{J}}$ is

$$\bar{\mathbf{J}} = \begin{bmatrix} & & 1 \\ & \iddots & \\ 1 & & \end{bmatrix}.$$

From (20), the coefficient vector for the maximum entropy prediction follows as:

$$\mathbf{f} = \mathbf{R}_{N-2}^{-1} \mathbf{r}_{N-1}, \quad (21)$$

and the next predicted correlation coefficient is

$$\zeta_{\max} = \mathbf{r}_{N-1}^H \bar{\mathbf{J}} \mathbf{R}_{N-2}^{-1} \mathbf{r}_{N-1}. \quad (22)$$

Thus, the maximum entropy solution can be expressed in two ways: either by substituting the value of $\zeta$ from (12) into (19), or directly via expression (21). The resulting spatial spectra are identical because both approaches define the same extension of the covariance matrix.

Let us examine the *minimum entropy* solution and explore alternative methods for obtaining the minimum entropy linear prediction vector beyond expression (19).

In contrast to the strictly positive definite maximum entropy extension, the minimum entropy extension $\mathbf{R}_N$ of the covariance matrix $\mathbf{R}_{N-1}$ is singular. At the first step of extending the covariance matrix, there are two variants of singular extension, which are defined in (13) and (14). Both options yield the same result since they generate equivalent spaces of the covariance matrix. Further singular extension of the matrix $\mathbf{R}_N$ is unique, as guaranteed by the infinite singular extension theorem [11]. In turn, the vector of linear prediction coefficients (19) and any variant of $\zeta$ (13) or (14) uniquely determine the covariance matrix extension and the corresponding minimum entropy spectrum.

The vector of the linear prediction coefficients is orthogonal to the space spanned by the rows of the covariance matrix $\mathbf{R}_N$,

$$\begin{bmatrix} r_0 & \mathbf{r}_N^H(\zeta_{\min}) \\ \mathbf{r}_\mathbf{N}(\zeta_{\min}) & \mathbf{R}_{N-1} \end{bmatrix} \begin{bmatrix} 1 \\ \mathbf{f} \end{bmatrix} = \begin{bmatrix} 0 \\ \mathbf{0} \end{bmatrix}. \quad (23)$$

Consequently, the minimum-entropy linear prediction vector is the eigenvector corresponding to the zero eigenvalue of the extended covariance matrix $\mathbf{R}_N(\zeta_{\min1,2})$.

Under the standard signal model, the covariance matrix $\mathbf{R}_{N-1}$ is full rank, whereas its minimum-entropy extension becomes singular. However, the minimum entropy criterion essentially assumes extrapolating signal wavefronts beyond the physical antenna aperture and does not require the original covariance matrix to be non-singular.

We consider a singular covariance matrix obtained by removing the uncorrelated noise component. The rank of the resulting matrix

$$\mathbf{C}_{N-1} = \mathbf{R}_{N-1} - \sigma^2\mathbf{I} = \mathbf{ASA}^H \tag{24}$$

equals the number of signal sources, rank($\mathbf{ASA}^H$) = $d$.

Let us derive a recurrence formula for the singular matrix extension that yields the minimum-entropy linear prediction coefficients. Assume the noise-free covariance matrix have rank $d<N$. It follows that any $d$+1 columns of $\mathbf{C}_{N-1}$ are linearly dependent. Thus, for the first $d$+1 columns, there exist coefficients $u_0, u_1, \ldots, u_d$, such that:

$$\begin{bmatrix} c_0 & c_1^* & \cdots & c_d^* \\ c_1 & c_0 & \cdots & c_{d-1}^* \\ \cdots & \cdots & \cdots & \cdots \\ c_{N-1} & c_{N-2} & \cdots & c_{N-d-1} \end{bmatrix} \begin{bmatrix} u_1 \\ u_2 \\ \cdots \\ u_{d+1} \end{bmatrix} = \begin{bmatrix} 0 \\ 0 \\ \cdots \\ 0 \end{bmatrix}. \tag{25}$$

These coefficients are proportional to the cofactors of the elements in the first row of determinant $D_d$ for matrix $\mathbf{C}_{N-1}$ [12]. Since appending a row preserves the matrix's rank, it follows that the first $d$+1 columns remain linearly dependent,

$$\begin{bmatrix} c_0 & c_1^* & \cdots & c_d^* \\ c_1 & c_0 & \cdots & c_{d-1}^* \\ \cdots & \cdots & \cdots & \cdots \\ c_{N-1} & c_{N-2} & \cdots & c_{N-d-1} \\ c_N & c_{N-1} & \cdots & c_{N-d} \end{bmatrix} \begin{bmatrix} u_1 \\ u_2 \\ \cdots \\ u_{d+1} \end{bmatrix} = \begin{bmatrix} 0 \\ 0 \\ \cdots \\ 0 \\ 0 \end{bmatrix}, \tag{26}$$

where the singular extension element, $\zeta$, is defined as

$$c_N = -\frac{u_2}{u_1}c_{N-1} - \frac{u_3}{u_1}c_{N-2} - \frac{u_3}{u_1}c_{N-2} - \cdots - \frac{u_{d+1}}{u_1}c_{N-d}. \tag{27}$$

This defines a recurrence relation that recursively updates the spatial correlation function and its corresponding Toeplitz covariance matrix, preserving their inherent rank. The coefficients $u_i$ are the cofactors of the first-row elements of matrix $\mathbf{C}$, though they can also be defined alternatively. Expression (25) implies that the forecast vector $\mathbf{u}$ is the right singular vector of the matrix formed by the first $d$+1 columns of $\mathbf{C}_{N-1}$, associated with the zero singular value. By complementing $\mathbf{u}$ with zeros to match the larger dimension, the resulting vector lies in the null space of $\mathbf{C}_{N-1}$, confirming it as an eigenvector of its zero eigenvalue:

$$\mathbf{C}_{N-1}\begin{bmatrix} \mathbf{u}_{d+1} \\ \mathbf{0}_{N-d-1} \end{bmatrix} = \mathbf{0}. \tag{28}$$

Furthermore, because adding $\sigma^2\mathbf{I}$ to $\mathbf{C}_{N-}$ simply shifts all its eigenvalues by $\sigma^2$ without altering its eigenvectors, we can express this as:

$$\mathbf{R}_{N-1}\begin{bmatrix} \mathbf{u}_{d+1} \\ \mathbf{0}_{N-d-1} \end{bmatrix} = \sigma^2\begin{bmatrix} \mathbf{u}_{d+1} \\ \mathbf{0}_{N-d-1} \end{bmatrix}. \tag{29}$$

Consequently, the zero-padded vector $\mathbf{u}$ acts as an eigenvector of the covariance matrix $\mathbf{R}_{N-1}$, with an associated eigenvalue equal to the noise variance $\sigma^2$.

The entire correlation sequence can be generated using $d$ initial values $(r_0, r_1, \ldots, r_d)$ and the noise eigenvector, $\mathbf{u}$, of the covariance matrix. While the Fourier transform of this infinite sequence yields the minimum entropy spatial spectrum, calculating it is unnecessary. Because the correlation sequence is recursive, its spectral characteristics are directly determined by the transfer function of a recursive linear filter represented by a polynomial:

$$H(z) = \frac{1}{u_1 + u_2 z^{-1} + u_3 z^{-2} + \cdots + u_{d+1} z^{-d}} = \frac{1}{B(z)}, \tag{30}$$

where $z = \exp\left(j2\pi\frac{\Delta}{\lambda}\sin(\theta)\right)$.

The spatial spectrum of signals is determined by the expression

$$P(\theta) = \frac{1}{\left|\sum_{k=1}^{d+1} u_k \exp(j2\pi(k-1)\frac{\Delta}{\lambda}\sin(\theta))\right|^2} = \frac{1}{\left|\mathbf{a}^H(\theta)\mathbf{u}\right|^2}. \tag{31}$$

The spectral estimate in (31) is derived from the minimum entropy criterion of the extrapolated correlation function. Physically, this equates to extending the plane wavefronts of the received signals beyond the boundary of the antenna aperture. Because vector $\mathbf{u}$ in (31) is an eigenvector corresponding to the noise eigenvalue of the covariance matrix, this estimate also represents the Pisarenko spatial spectrum [4, 13]. Therefore, the Pisarenko method extrapolates the correlation function beyond the observed region, much like the maximum entropy method. Unlike the latter, it relies on minimizing the entropy of the process's power spectral density. This corresponds to extrapolating the plane wavefronts of the sources beyond the antenna aperture.

## III. Multi-Channel Linear Prediction

We define multi-channel linear prediction as a process where each prediction step extrapolates an antenna subarray rather than a single element. We demonstrate that the ESPRIT algorithm is fundamentally equivalent to this multi-channel linear prediction approach.

We begin with the classical formulation of the ESPRIT method, which we then reformulate within a linear prediction framework to interpret it as an extrapolation of source wavefronts beyond the antenna aperture. A distinctive feature of ESPRIT and related subspace methods [6, 17, 18, 19, 21, 23] is that they require neither precise sensor calibration nor prior knowledge of the global array geometry. Instead, they rely on an array geometry composed of matched sensor pairs, where each pair shares an identical spatial displacement vector. The DOA estimates are then extracted directly from the generalized eigenvalues of a constructed matrix pencil.

In ESPRIT-like methods, it is convenient to consider the antenna array as consisting of two identical subarrays spatially displaced by the vector $\Delta$. The antenna array receives narrowband signals with plane wavefronts arriving from external sources as presented in Fig. 2. Let $\mathbf{a}_m = \mathbf{a}(\theta_m)$ denote the steering vector of the $m$-th source at the first subarray, corresponding to the angular coordinate $\theta_m$. Because the subarrays are identical, the signals they receive from the same sources differ only by progressive phase shifts $\varphi_1,..., \varphi_d$. The output signals of the first and second subarrays are given by

$$\mathbf{X}_1(t) = [\mathbf{a}(\theta_1), \dots, \mathbf{a}(\theta_d)]\mathbf{s}(t) + \mathbf{n}_1(t) = \mathbf{A}\mathbf{s}(t) + \mathbf{n}_1(t), \quad (32)$$

$$\mathbf{X}_2(t) = [\mathbf{a}(\theta_1)e^{-j\varphi_1}, \dots, \mathbf{a}(\theta_d)\mathbf{a}e^{-j\varphi_d}]\mathbf{s}(t) + \mathbf{n}_2(t) = \mathbf{A\Phi s}(t) + \mathbf{n}_2(t), \quad (33)$$

where $\mathbf{n}_1(t)$, $\mathbf{n}_2(t)$ are the additive noise vectors of the first and second subarrays; $\mathbf{\Phi} = \mathrm{diag}[\exp\{j\varphi_1\},..., \exp\{j\varphi_d\}]$, $\varphi_m = 2|\Delta| \sin\theta_m$.

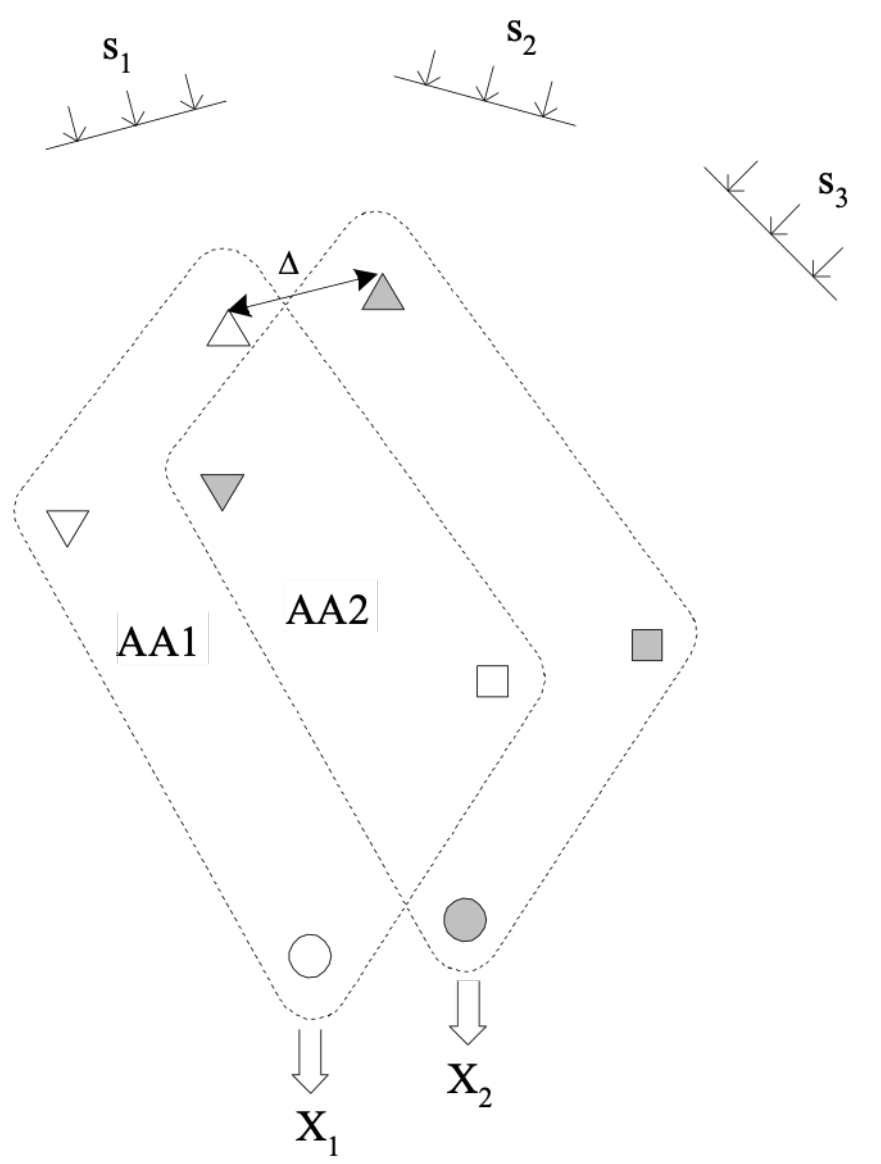

Fig. 2. ESPRIT antenna array layout.

In generalized notation, the signals from the antenna array are expressed as follows:

$$\mathbf{X}(t) = \begin{bmatrix} \mathbf{X}_1(t) \\ \mathbf{X}_2(t) \end{bmatrix} = \begin{bmatrix} \mathbf{A} \\ \mathbf{A\Phi} \end{bmatrix} \mathbf{s}(t) + \mathbf{n}(t) = \mathbf{Gs}(t) + \mathbf{n}(t). \quad (34)$$

The source signals $s(t)$ are zero-mean, temporary uncorrelated, complex Gaussian processes with a covariance matrix $\mathbf{S} = \mathrm{E}[\mathbf{s}(t)\mathbf{s}^H(t)]$. The noise $\mathbf{n}(t)$ also is modeled as a zero-mean, spatially and temporally white Gaussian process with variance $\sigma^2$. It is assumed to be uncorrelated with the source signals. The covariance matrix of the signals received by the antenna array is given by:

$$\mathbf{R} = E[\mathbf{X}(t)\mathbf{X}^H(t)] = \mathbf{GSG}^H + \sigma^2\mathbf{I}. \quad (35)$$

We assume that matrices $\mathbf{G}$ and $\mathbf{S}$ have full rank $\mathrm{rank}(\mathbf{G}) = \mathrm{rank}(\mathbf{S}) = d$, implying that the source signals are not fully coherent.

In the ESPRIT method, the directions of arrival $\theta_m$ are determined from the non-zero generalized eigenvalues $\lambda_m$ ($m = 1, ..., d$) of the matrix pencil $\{\mathbf{C}_{11}, \mathbf{R}_{12}\}$, which correspond to the non-zero solutions of the equation

$$\det[\mathbf{C}_{11} - \lambda\mathbf{R}_{12}] = 0, \quad (36)$$

$$\theta_m = \arcsin\left(\frac{\lambda_m}{2\pi|\Delta|}\right), \quad (37)$$

where $\mathbf{C}_{11} = \mathbf{R}_{11} - \sigma^2\mathbf{I}$ is the covariance matrix of the first subarray without noise and $\mathbf{R}_{12}$ is the cross-covariance matrix of the first and second subarrays:

$$\mathbf{R} = \begin{bmatrix} \mathbf{R}_{11} & \mathbf{R}_{12} \\ \mathbf{R}_{21} & \mathbf{R}_{22} \end{bmatrix}, \quad (38)$$

$$\mathbf{C} = \mathbf{R} - \sigma^2\mathbf{I}_{2m} = \begin{bmatrix} \mathbf{C}_{11} & \mathbf{R}_{12} \\ \mathbf{R}_{21} & \mathbf{C}_{22} \end{bmatrix} = \begin{bmatrix} \mathbf{ASA}^H & \mathbf{AS\Phi}^H\mathbf{A}^H \\ \mathbf{A\Phi SA}^H & \mathbf{A\Phi S\Phi}^H\mathbf{A}^H \end{bmatrix}. \quad (39)$$

We will now consider the ESPRIT method from the perspective of extrapolating the correlation function beyond the physical antenna array.

The antenna system described above — consisting of two identical subarrays spatially shifted by vector $\Delta$ — will be treated here as a modular array formed by two identical modules with shifted phase centers. First, we predict the cross-covariance matrix for the second and third (virtual) modules of the antenna array presented in Fig. 3. Subsequent predictions are then generated recursively.

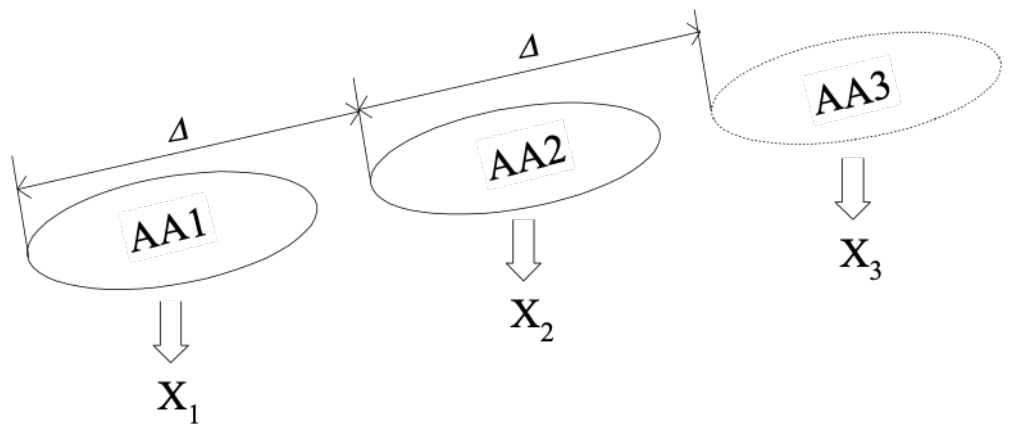

Fig. 3. Layout of the antenna subarrays.

Extrapolating the wavefronts of signal sources in a modular antenna array is equivalent to enforcing the minimum entropy criterion. While a single step of uniform linear array (ULA) extrapolation predicts the signal or correlation function for a single virtual element, extrapolation in a modular array predicts either the signals across $N$ spatial points or the cross-covariance matrix between the predicted and preceding subarrays.

Assuming spatial stationarity, the noise-free covariance matrix of the antenna array outputs is block-Toeplitz:

$$\mathbf{C} = \mathbf{R} - \sigma^2\mathbf{I}_{2m} = \begin{bmatrix} \mathbf{C}_1 & \mathbf{C}_2 \\ \mathbf{C}_2^H & \mathbf{C}_1 \end{bmatrix}. \quad (40)$$

Unlike conventional linear prediction which uses scalar-element vectors, multichannel linear prediction for wavefronts outside an antenna array utilizes matrix-element prediction vectors. This shift from scalars to matrices introduces additional degrees of freedom, which can be used to develop new methods for estimating matrix parameters and source angular coordinates under the minimum entropy criterion. We first introduce a block prediction matrix configuration to highlight the core principles of minimum-entropy multichannel prediction, and then demonstrate that ESPRIT estimates can be derived as a special case of this proposed model.

We now introduce the block matrix $\mathbf{B}$,

$$\mathbf{B} = \begin{bmatrix} \mathbf{B}_1 \\ \mathbf{B}_2 \end{bmatrix},$$

where $\mathbf{B}_1$, $\mathbf{B}_2$ are $N \times N$ matrices satisfy the condition

$$\mathbf{CB} = \begin{bmatrix} \mathbf{C}_1 & \mathbf{C}_2 \\ \mathbf{C}_2^H & \mathbf{C}_1 \end{bmatrix} \begin{bmatrix} \mathbf{B}_1 \\ \mathbf{B}_2 \end{bmatrix} = \begin{bmatrix} \mathbf{0} \\ \mathbf{0} \end{bmatrix}. \quad (41)$$

Note that rank($\mathbf{C}_1$) = rank($\mathbf{C}_2$) = $d$.

The block matrix $\mathbf{B}$ can be viewed as a generalized eigenvector of the matrix $\mathbf{C}$ corresponding to a zero eigenvalue. Because

$$\dim(\ker(\mathbf{C})) = 2N - d > N, \quad (42)$$

these vectors are not unique, which allows for some flexibility in its estimation.

On the other hand, equation (36) defines the generalized eigenvalues and eigenvectors of the matrix pencil $\{\mathbf{C}_{11}, \mathbf{R}_{12}\}$ (denoted here as $\{\mathbf{C}_1, \mathbf{C}_2\}$). This can be expressed as

$$(\mathbf{C}_1 - \lambda \mathbf{C}_2)\mathbf{u} = \mathbf{0}, \quad (43)$$

where $\lambda$ and $\mathbf{u}$ are the generalized eigenvalue and corresponding generalized eigenvector, respectively. Equation (43) can be rewritten as follows:

$$[\mathbf{C}_1 \quad \mathbf{C}_2] \begin{bmatrix} \mathbf{u} \\ -\lambda \mathbf{u} \end{bmatrix} = 0. \quad (44)$$

A vector $\mathbf{u}$ satisfying (43) or (44) is a generalized eigenvector of the pencil $\{\mathbf{C}_1, \mathbf{C}_2\}$, meaning the vector $[\mathbf{u}^T, -\lambda\mathbf{u}^T]^T$ is both a right singular vector of $[\mathbf{C}_1, \mathbf{C}_2]$ (zero singular value) and an eigenvector of $\mathbf{C}$ (zero eigenvalue), thus satisfying condition (41) for the columns of $\mathbf{B}$. The block matrix $\mathbf{B}$ can thus be defined as

$$\mathbf{B}_1 = \mathbf{U}, \quad \mathbf{B}_2 = -\mathbf{U\Lambda}, \quad \mathbf{B} = \begin{bmatrix} \mathbf{U} \\ -\mathbf{U\Lambda} \end{bmatrix}, \quad (45)$$

where $\mathbf{U} = [\mathbf{u}_1, \mathbf{u}_2, ..., \mathbf{u}_N]$ and $\mathbf{\Lambda} = \mathrm{diag}\{\lambda_m\}$. Here, $\mathbf{u}_m$ and $\lambda_m$ (for $m$=1, ..., $N$) denote the $m$-th generalized eigenvector and the corresponding generalized eigenvalue of the pencil $\{\mathbf{C}_1, \mathbf{C}_2\}$.

Order the eigenvalues $\lambda_m$ so that $\lambda_{d+1} = \lambda_{d+2} = \ldots = \lambda_N = 0$. Consequently, the vectors $\mathbf{u}_{d+1}, \mathbf{u}_{d+2}, ..., \mathbf{u}_N$ are also eigenvectors of matrix $\mathbf{C}_1$ corresponding to its zero eigenvalues, while $\mathbf{u}_1, \mathbf{u}_2, ..., \mathbf{u}_d$ are proportional to the columns of the matrix $\mathbf{A}(\mathbf{A}^H\mathbf{A})^{-1}\mathbf{\Phi}^H$.

Therefore, matrix $\mathbf{B}$ can be also expressed as:

$$\mathbf{B} = \begin{bmatrix} \mathbf{A}(\mathbf{A}^H\mathbf{A})^{-1} & \mathbf{U}_{d+1...N} \\ \mathbf{A}(\mathbf{A}^H\mathbf{A})^{-1}\mathbf{\Phi}^H & \mathbf{0} \end{bmatrix}, \quad (46)$$

where $\mathbf{U}_{d+1...N}$ is a matrix consisting of the eigenvectors associated with the zero eigenvalues of the subarray covariance matrix. Because matrix $\mathbf{B}$ is not uniquely defined, we present one possible definition to demonstrate the existence of a matrix that satisfies the above conditions. Practical methods for estimating $\mathbf{B}$ are discussed below.

The block matrix $\mathbf{B}$ defined in (45) acts as the multichannel linear predictor for the subarray, shifted by the vector $\Delta$. Furthermore, the cross-covariance matrix $\mathbf{C}_3$ between the second and third subarrays is defined by the following expression:

$$\mathbf{C}_3\mathbf{U\Lambda} = \mathbf{C}_2\mathbf{U}, \quad (47)$$

and for an arbitrary ($k$+1)-th step of predicting, respectively

$$\mathbf{C}_{k+1}\mathbf{U\Lambda} = \mathbf{C}_k\mathbf{U}. \quad (48)$$

Equation (48) is a multichannel autoregressive model, extending the standard autoregressive formulation to account for multiple channels.

For this antenna array configuration, the steering vector corresponding to an incoming signal from direction $\theta$ is structured as follows:

$$\mathbf{a}(\theta) = \begin{bmatrix} \mathbf{a}_1(\theta) \\ \gamma \mathbf{a}_1(\theta) \end{bmatrix}, \quad (49)$$

where $\mathbf{a}_1(\theta)$ is the amplitude-phase distribution vector of the reference signal received from direction $\theta$ at the aperture of the first subarray, and $\gamma = \exp(j2\pi|\Delta| \sin\theta)$.

Following the scalar case, the multichannel spatial spectrum is determined by its block-matrix prediction filter:

$$P(\theta) = \frac{1}{\|\mathbf{a}(\theta)^H\mathbf{B}\|^2}. \quad (50)$$

Without specifying the structure of matrix $\mathbf{B}$, and relying solely on the fact that its columns span the noise subspace of the covariance matrix $\mathbf{R}$ (or the null space of matrix $\mathbf{C}$), we can find that expression (50) represents the MUSIC spatial spectrum estimate. However, by exploiting the specific structure of the eigenvector matrix $\mathbf{B}$, the extremum condition for functional (50) can be reformulated as

$$\mathbf{a}^H(\theta)\mathbf{B} = \mathbf{0}, \quad (51)$$

from where, considering (45) and (49), we obtain

$$\mathbf{a}^H(\theta)\mathbf{U} = \gamma \mathbf{a}^H(\theta)\mathbf{U\Lambda}, \quad (52)$$

$$\gamma_m = \lambda_m^* = \exp(j2\pi|\Delta|\sin(\theta_m)),\ m = 1, ..., d.$$

Thus, the ESPRIT spatial spectrum estimate can be interpreted as a multichannel autoregressive prediction of signals beyond the antenna aperture based on the minimum entropy criterion. Unlike the single-channel minimum entropy prediction in Pisarenko case, ESPRIT predicts the plane wavefronts of the sources across the aperture of the next $N$-element subarray.

Similar to the single-channel prediction, the multi-channel prediction can be optimized using not only the minimum entropy criterion, but also the maximum entropy criterion. Below we derive the maximum entropy spatial spectrum estimate for multichannel subarray prediction.

The signal and covariance matrix models for the antenna array remain unchanged, corresponding to (34) and (35). In the single-channel case, the maximum entropy solution reduces to a single row (or column) of the signal covariance matrix. For the multichannel case, this scalar covariance structure is extended, with the standard covariance matrix replaced by a corresponding block matrix:

$$\begin{bmatrix}\mathbf{R}_{11} & \mathbf{R}_{12}\\ \mathbf{R}_{12}^{H} & \mathbf{R}_{11}\end{bmatrix}^{-1} = \begin{bmatrix}\mathbf{K}^{-1} & \mathbf{K}^{-1}\mathbf{R}_{12}\mathbf{R}_{11}^{-1}\\ -\mathbf{R}_{11}^{-1}\mathbf{R}_{12}^{H}\mathbf{K}^{-1} & \mathbf{R}_{11}^{-1}+\mathbf{R}_{11}^{-1}\mathbf{R}_{12}^{H}\mathbf{K}^{-1}\mathbf{R}_{12}\mathbf{R}_{11}^{-1}\end{bmatrix}, \quad (53)$$

where

$$\mathbf{K} = \mathbf{R}_{11} - \mathbf{R}_{11}^{-1}\mathbf{R}_{12}^{H}.$$

Matrix $\mathbf{K}$ normalizes the first column of the inverse block matrix (53) so that its first block becomes the identity matrix. Right-multiplying the first block column of (53) by $\mathbf{K}$ yields the block vector of the multichannel autoregressive maximum entropy prediction:

$$\mathbf{F} = \begin{bmatrix}\mathbf{I}\\ -\mathbf{R}_{11}^{-1}\mathbf{R}_{12}^{H}\end{bmatrix}. \quad (54)$$

The multichannel estimate of the maximum entropy spatial spectrum is given by:

$$P_{ESPME}(\theta) = \frac{1}{\|\mathbf{a}(\theta)^{H}\mathbf{F}\|^{2}} = \frac{1}{\|\mathbf{a}_1(\theta)^{H} - \mathbf{a}_2(\theta)^{H}\mathbf{R}_{11}^{-1}\mathbf{R}_{12}^{H}\|^{2}}, \quad (55)$$

where $\mathbf{a}_1(\theta)$ и $\mathbf{a}_2(\theta)$ are the steering vectors for the first and second subarrays, respectively.

The angular coordinate estimates derived from the spatial spectrum (55) correspond to the zeros of the function

$$\|\mathbf{a}_1(\theta)^{H} - \mathbf{a}_2(\theta)^{H}\mathbf{R}_{11}^{-1}\mathbf{R}_{12}^{H}\|^{2} = 0. \quad (56)$$

Due to the specific configuration of the antenna array, the amplitude-phase distribution vectors are collinear, i.e., $\mathbf{a}_1(\theta) = \lambda\mathbf{a}_2(\theta)$, where $\lambda = \exp(-j2\pi|\Delta|\sin\theta)$. Consequently, equation (56) simplifies to:

$$\mathbf{R}_{12}\mathbf{R}_{11}^{-1}\mathbf{a}_1(\theta) = \lambda\mathbf{a}_1(\theta). \quad (57)$$

The equation provides exact solutions when the correlation function of the received signals follows a maximum-entropy autoregressive model. Specifically, these solutions are the eigenvalues of the matrix $\mathbf{R}_{12}\mathbf{R}_{11}^{-1}$, which define the angular coordinates of the signal sources.

Consequently, the approximate ESPRIT solutions are obtained by solving the standard eigenvalue problem for the matrix $\mathbf{R}_{12}\mathbf{R}_{11}^{-1}$. These solutions effectively predict the signal wavefronts beyond the physical antenna aperture while maximizing the entropy of the block correlation function.

## IV. Multi-Channel Linear Prediction of Non-Gaussian Signals

The Virtual-ESPRIT Algorithm (VESPA) [7, 19, 20] extends matrix pencil methods by utilizing fourth-order cumulants for non-Gaussian signals. Like standard ESPRIT, VESPA extracts source angles from the generalized eigenvalues of a matrix pencil. However, while standard approaches require a second physical antenna array to construct a cross-covariance matrix, VESPA replaces it with a virtual array. Cross-correlations are estimated directly from the cumulants of the single physical array, eliminating the need for additional subarray.

In this section, we reframe VESPA-type methods. We transition from their classical, cumulant-matrix definition to their interpretation as linear signal prediction beyond the physical antenna aperture.

Unlike the original ESPRIT method, VESPA uses a single $N$-element antenna array of arbitrary configuration presented in Fig. 4. Two channels of this array must be identical and spatially shifted by a known vector $\Delta$.

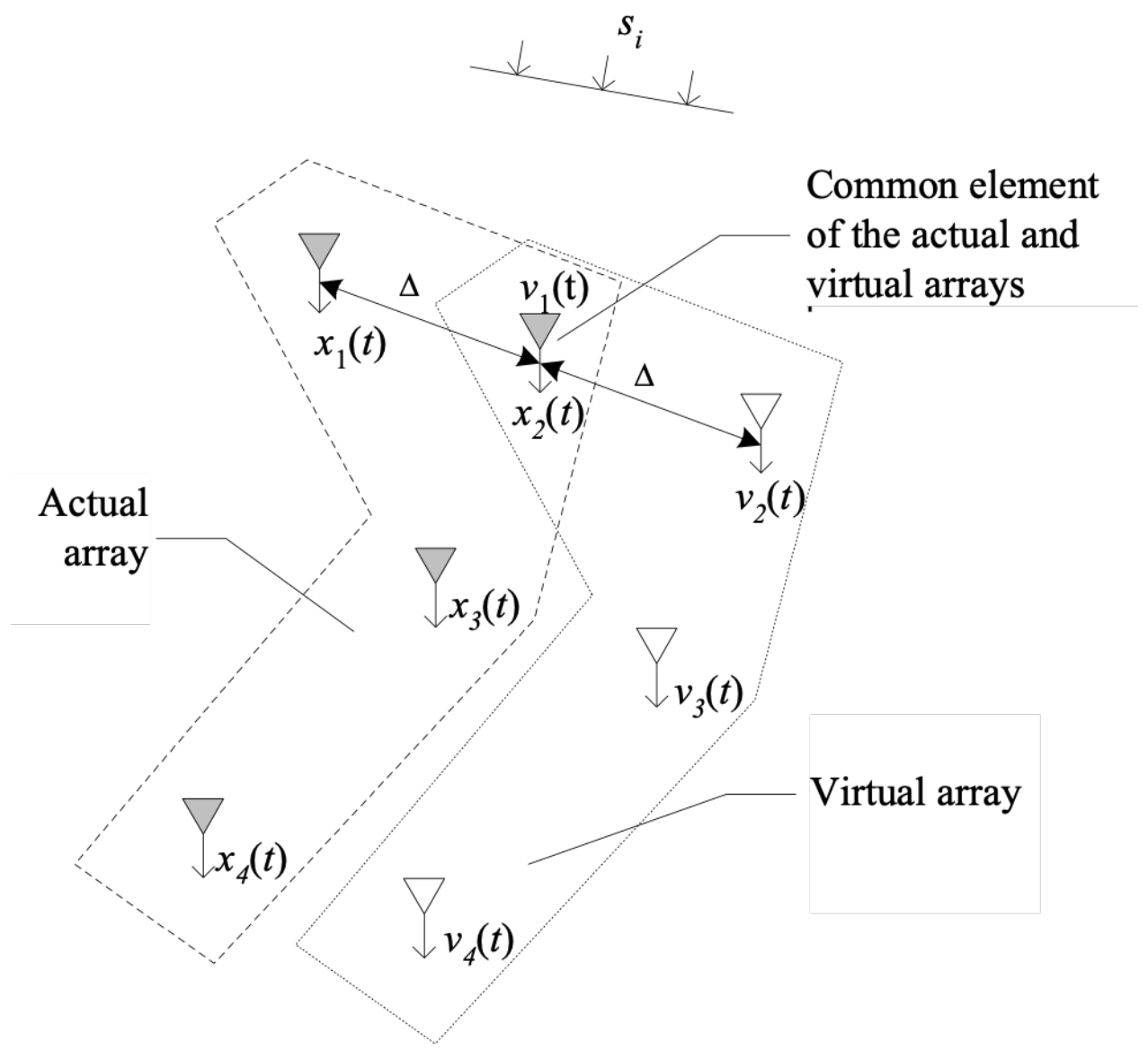

Fig. 4. Actual and virtual arrays in the VESPA framework.

Calculating the angular coordinates of the signal sources using the matrix pencil method requires estimating both the spatial covariance matrix of the main antenna array and the cross-covariance matrix between the main array and its spatial shifted version by the vector $\Delta$. This shifted copy is physically absent, serving instead as the virtual array illustrated in Fig. 4.

Cumulant matrix of the physical antenna array for non-Gaussian sources takes the following form [7]:

$$\tilde{\mathbf{R}}_{XX} = \mathbf{A}\boldsymbol{\Gamma}\mathbf{A}^{H}, \quad (58)$$

where $\mathbf{A}$ is the array steering matrix of the signal wavefronts on the real antenna aperture and $\boldsymbol{\Gamma}$ is the diagonal matrix of the source signals' fourth-order cumulants, $\boldsymbol{\Gamma} = \mathrm{diag}\left\{\gamma_k^{(4)}\right\}$.

The mutual fourth-order cumulant matrix of the real and virtual antenna arrays is defined as follows:

$$\tilde{\mathbf{R}}_{XV} = \sum_{k=1}^{d}\gamma_k^{(4)}e^{j\varphi_k}\mathbf{a}_k\mathbf{a}_k^{H} = \mathbf{A}\boldsymbol{\Gamma}\boldsymbol{\Phi}^{H}\mathbf{A}^{H}, \quad (59)$$

where $\varphi_k = 2\pi|\Delta|\sin\theta_k$ and $\theta_k$ is the direction of arrival of the $k$-th source signal.

By analyzing the structure of cumulant matrices (65) and (69), we show that the generalized eigenvalues of the matrix pencil $\{\tilde{\mathbf{R}}_{XX}, \tilde{\mathbf{R}}_{XV}\}$ directly yield the values $\xi_k = e^{j\varphi_k}$ corresponding to the signal bearings. This behavior mirrors the $\{\mathbf{C}_{XX}, \mathbf{R}_{XV}\}$ pencil that the ESPRIT method would employ if a second antenna array existed. This can be verified through the pencil equation. Since

$$\widetilde{\mathbf{R}}_{XX} - \xi \widetilde{\mathbf{R}}_{XV} = \mathbf{A\Gamma A}^H - \xi\ \mathbf{A\Gamma\Phi}^H\mathbf{A}^H$$

$$= \mathbf{A}(\mathbf{\Gamma A}^H - \xi \mathbf{A\Gamma\Phi}^H)\mathbf{A}^H, \qquad (60)$$

and both $\mathbf{\Gamma}$ and $\mathbf{\Phi}$ are diagonal matrices, the rank of the matrix $\widetilde{\mathbf{R}}_{XX} - \xi \widetilde{\mathbf{R}}_{XV}$ jumps precisely when the parameters $\xi_k$ equal the diagonal elements of $\mathbf{\Phi}$.

Thus, higher-order statistics enable the prediction of source wavefronts on a virtual aperture without constraining the antenna array configuration. Unlike the ESPRIT method, which requires $N$ pairs of identical sensors, only a single identical pair is necessary here. Furthermore, the inherent noise-filtering properties of fourth-order cumulants eliminate the need to pre-process the covariance matrix to remove spatially uncorrelated noise. This suppression applies to both white and time-correlated noise. In practice, however, effective noise reduction depends on averaging a sufficient number of signal samples.

We employ higher-order statistics to predict the field distribution across the aperture of a second identical antenna array. This represents the first step in extending the plane wavefronts of signal sources beyond the existing antenna aperture. As a result, we obtain two cumulant matrices, $\widetilde{\mathbf{R}}_{XX}$ and $\widetilde{\mathbf{R}}_{XV}$. These matrices can be represented as sums of dyadic wavefront matrices originating from sources at two identical, spatially offset apertures. Our objective is to extend the source wavefronts further while maintaining a fixed rank for the cumulant matrices. Since matrices derived from fourth-order cumulants differ structurally from conventional covariance matrices, we analyze this process in detail.

The full cumulant matrix of the signals received by the antenna array, which comprises both real and virtual subarrays, takes the form:

$$\widetilde{\mathbf{R}} = \begin{bmatrix} \widetilde{\mathbf{R}}_{XX} & \widetilde{\mathbf{R}}_{XV} \\ \mathbf{R}_{XV}^H & \widetilde{\mathbf{R}}_{XX} \end{bmatrix}. \qquad (61)$$

The matrix has a block-Toeplitz structure and can be defined directly by the source wavefronts:

$$\widetilde{\mathbf{R}} = \begin{bmatrix} \mathbf{A\Gamma A}^H & \mathbf{A\Gamma\Phi}^H\mathbf{A}^H \\ \mathbf{A\Phi\Gamma A}^H & \mathbf{A\Gamma A}^H \end{bmatrix}. \qquad (62)$$

Note that the complete antenna array's cumulant matrix adopts form (62) only when all submatrices are estimated using fourth-order cumulants. If the matrix of the real antenna array is replaced by second-order statistics—which are available for the real array but not for the virtual one—then the generalized eigenvalues of the pencil $\{\widetilde{\mathbf{R}}_{XX}, \widetilde{\mathbf{R}}_{XV}\}$ will no longer determine the bearings of the signal sources. Consequently, it becomes impossible to extrapolate the sources' plane wavefronts beyond the real and virtual arrays using this matrix combination.

Formula (5) is inapplicable because it relies on the assumption of Gaussian statistics. For non-Gaussian signals, maximizing or minimizing process entropy is no longer equivalent to optimizing the determinant of the covariance matrix. Consequently, to extend the signal beyond the antenna aperture, we return to the criterion for plane wavefront extension with a fixed number of non-Gaussian sources (the cumulant matrix rank). Formally, this requires finding a block-Toeplitz extension of the cumulant matrix (61) that retains the rank of the original matrix.

By analogy with the extension of covariance matrices, we introduce the block matrix

$$\mathbf{B} = \begin{bmatrix} \mathbf{B}_1 \\ \mathbf{B}_2 \end{bmatrix},$$

whose blocks are square matrices $\mathbf{B}_1$ and $\mathbf{B}_2$ of dimension $N \times N$, such that

$$\widetilde{\mathbf{R}}\mathbf{B} = \begin{bmatrix} \widetilde{\mathbf{R}}_{XX} & \widetilde{\mathbf{R}}_{XV} \\ \mathbf{R}_{XV}^H & \widetilde{\mathbf{R}}_{XX} \end{bmatrix} \begin{bmatrix} \mathbf{B}_1 \\ \mathbf{B}_2 \end{bmatrix} = \begin{bmatrix} \mathbf{0} \\ \mathbf{0} \end{bmatrix}. \qquad (63)$$

Since the projected cumulant matrix $\widetilde{\mathbf{R}}_\infty$ is block Toeplitz and its rank is fixed, $\mathrm{rank}(\widetilde{\mathbf{R}}_\infty) = \mathrm{rank}(\widetilde{\mathbf{R}}) = \mathrm{rank}(\widetilde{\mathbf{R}}_{XX}) = d$, the block eigenvector of the cumulant matrix $\widetilde{\mathbf{R}}$ uniquely determines all block eigenvectors spanning the null space of the extended cumulant matrix, which can be expressed as follows:

$$\begin{bmatrix} \widetilde{\mathbf{R}}_{XX} & \widetilde{\mathbf{R}}_{XV} & \widetilde{\mathbf{R}}_{XZ} & \dots \\ \mathbf{R}_{XV}^H & \widetilde{\mathbf{R}}_{XX} & \widetilde{\mathbf{R}}_{XV} & \dots \\ \mathbf{R}_{XZ}^H & \mathbf{R}_{XV}^H & \widetilde{\mathbf{R}}_{XX} & \dots \\ \dots & \dots & \dots & \dots \end{bmatrix} \begin{bmatrix} \mathbf{B}_1 & \mathbf{0} & \mathbf{0} & \dots \\ \mathbf{B}_2 & \mathbf{B}_1 & \mathbf{0} & \dots \\ \mathbf{0} & \mathbf{B}_2 & \mathbf{B}_1 & \dots \\ \dots & \dots & \dots & \dots \end{bmatrix} = \begin{bmatrix} \mathbf{0} \\ \mathbf{0} \\ \mathbf{0} \\ \dots \end{bmatrix}, \qquad (64)$$

where $\widetilde{\mathbf{R}}_{XZ}$ denotes the subsequent predicted mutual cumulant matrix between the physical antenna array $\mathbf{X}$ and the predicted virtual array $\mathbf{Z}$. The antenna array configuration is presented in Fig. 5.

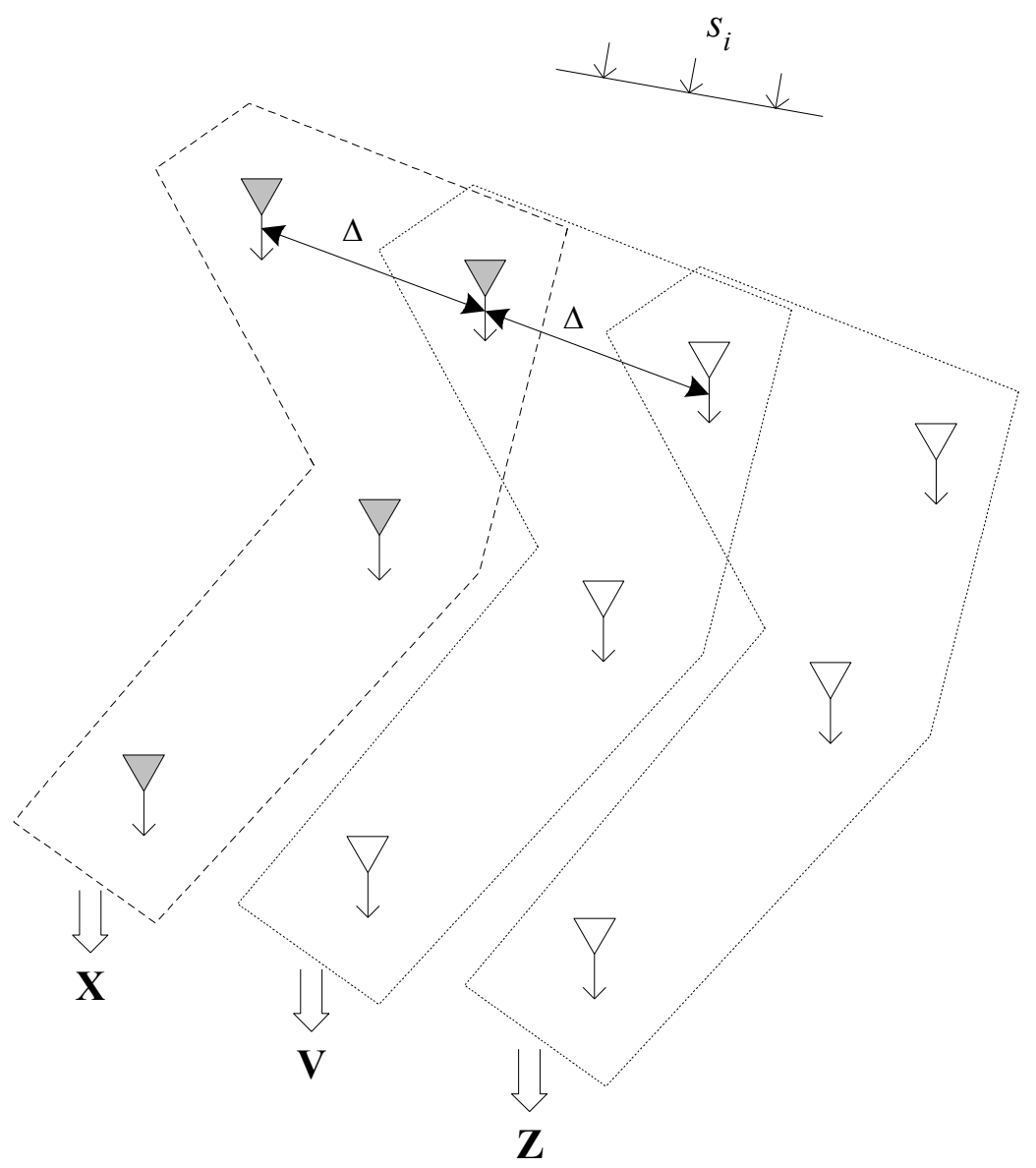

Fig. 5. Prediction virtual subarrays as copies of the actual array shifted by the vector $\Delta$

The validity of (64) can be demonstrated by isolating any $2N \times 2N$ diagonal block in $\widetilde{\mathbf{R}}_\infty$, which corresponds to two adjacent virtual antenna arrays. Because $\widetilde{\mathbf{R}}_\infty$ is block Toeplitz, this diagonal submatrix is identical to $\widetilde{\mathbf{R}}$ and thus shares its block eigenvector, $\mathbf{B}$. Since the rank of $\widetilde{\mathbf{R}}_\infty$ does not exceed that of $\widetilde{\mathbf{R}}$, any submatrix row above or below this block is a linear combination of $\widetilde{\mathbf{R}}$. Consequently, $\mathbf{B}$, padded with appropriately dimensioned zero blocks, forms a block eigenvector of $\widetilde{\mathbf{R}}_\infty$ associated with its zero eigenvalue.

Expression (64) shows that the block vector $\mathbf{B}$ acts as a predictor for the cumulant matrices of the real and virtual subarrays. Specifically, the cumulant matrix $\widetilde{\mathbf{R}}_{XZ}$ can be expressed as the solution to:

$$\widetilde{\mathbf{R}}_{XV}\mathbf{B}_1 = -\widetilde{\mathbf{R}}_{XZ}\mathbf{B}_2. \tag{65}$$

Alternatively, the matrix pencil $\{\widetilde{\mathbf{R}}_{XX}, \widetilde{\mathbf{R}}_{XV}\}$ can be expressed in the form

$$[\widetilde{\mathbf{R}}_{XX} \quad \widetilde{\mathbf{R}}_{XV}]\begin{bmatrix}\mathbf{u}\\ -\xi\mathbf{u}\end{bmatrix} = \mathbf{0}, \tag{66}$$

where $\mathbf{u}$ is the generalized eigenvector of the pencil, and $\xi$ is the generalized eigenvalue.

Comparing (63) and (66) shows the vectors $[\mathbf{u}^T, -\xi\mathbf{u}^T]^T$ in (66) lie in the column space of $\mathbf{B}$. Following the logic of the matrix pencil method, we can determine the components of the block matrix $\mathbf{B}$ as follows:

$$\mathbf{B}_1 = \mathbf{U};\ \mathbf{B}_2 = -\mathbf{U}\Xi, \tag{67}$$

where $\mathbf{U} = [\mathbf{u}_1, \mathbf{u}_2, \ldots, \mathbf{u}_N]$ and $\Xi = \mathrm{diag}\{\xi_k\}_{k=1,\ldots,N}$.

To confirm the existence of a block vector $\mathbf{B}$ in the form (67), we provide an example for the case where the true cumulant matrix has the form (62). Of the $N$ generalized eigenvalues of the pencil $\{\widetilde{\mathbf{R}}_{XX}, \widetilde{\mathbf{R}}_{XV}\}$, exactly $(N-d)$values of $\xi_k$ are zero, while $\xi_{d+1}, \ldots, \xi_N$ are assumed to be nonzero. Consequently, the eigenvectors $\mathbf{u}_{d+1}, \mathbf{u}_{d+2}, \ldots, \mathbf{u}_N$ span the null space of $\widetilde{\mathbf{R}}_{XX}$. The remaining eigenvectors $\mathbf{u}_1, \mathbf{u}_2, \ldots, \mathbf{u}_d$ lie in the column space of $\mathbf{A}(\mathbf{A}^H\mathbf{A})^{-1}\mathbf{\Phi}^H$. This demonstrates that the block vector $\mathbf{B}$ can take the form (46), consistent with the ESPRIT method.

The predicted cross-cumulant matrix of the $\mathbf{X}$ and $\mathbf{Z}$ subarrays is defined by the following equation:

$$\widetilde{\mathbf{R}}_{XZ}\Xi = \widetilde{\mathbf{R}}_{XV}\mathbf{U}. \tag{68}$$

At an arbitrary step ($k$+1), the cumulant matrix for the ($k$+1)-th antenna array satisfies the equation:

$$\widetilde{\mathbf{R}}_{k+1}\mathbf{U}\Xi = \widetilde{\mathbf{R}}_k\mathbf{U}. \tag{69}$$

Equation (68) defines a zero-residual multichannel autoregressive model with matrix coefficients $\mathbf{B}_1 = \mathbf{U}$ and $\mathbf{B}_2 = -\mathbf{U}\Xi$. Consequently, the VESPA virtual aperture expansion method can be interpreted as an extrapolation of plane wavefronts beyond both the real and the virtual subarrays. This extrapolation is driven by a block multichannel, minimum-entropy autoregressive process.

## V. Averaging Linear Prediction Estimates

Previous sections established that the Pisarenko, ESPRIT, and VESPA algorithms can be interpreted as techniques for extrapolating the signal correlation function or wavefronts beyond the antenna aperture. Yet, certain angular super-resolution methods—such as Capon [3] and MUSIC [5] (and its variants [18, 22-26])—do not inherently extend a single correlation function beyond the observed domain. We will demonstrate that these techniques can still be considered as liner prediction methods, provided the corresponding spatial spectrum estimates are averaged over subarrays.

The following expressions define the spatial spectrum estimates for the Capon and maximum entropy methods, respectively:

$$P_{CP}(\theta) = \frac{1}{\mathbf{a}(\theta)^H\mathbf{R}^{-1}\mathbf{a}(\theta)}, \tag{70}$$

$$P_{ME}(\theta) = \frac{\mathbf{1}_1^T\mathbf{R}^{-1}\mathbf{1}_1}{|\mathbf{a}(\theta)^H\mathbf{f}|^2}, \tag{71}$$

where $\mathbf{R}$ is the covariance matrix of the received signals, $\mathbf{f}$ is the vector of autoregressive parameters satisfying the maximum entropy criterion, $\mathbf{a}(\theta)$ is the steering vector, and $\mathbf{1}_1 = [1, 0, \ldots, 0]^T$.

The vector of autoregressive parameters, $\mathbf{f}$, represents the first column of the inverse covariance matrix $\mathbf{R}^{-1}$. Therefore, expression (71) can be rewritten as:

$$P_{ME}(\theta) = \frac{\mathbf{1}_1^T\mathbf{R}^{-1}\mathbf{1}_1}{|\mathbf{a}(\theta)^H\mathbf{R}^{-1}\mathbf{1}_1|^2}. \tag{72}$$

To demonstrate the relationship between the Capon method and maximum entropy autoregressive prediction, we use a triangular factorization of the inverse covariance matrix,

$$\mathbf{R}^{-1} = \mathbf{L}\mathbf{L}^H, \tag{73}$$

where $\mathbf{L}$ is a lower triangular matrix.

The Capon and maximum entropy spectral estimates are then given by the following expressions:

$$P_{CP}(\theta) = \frac{1}{\mathbf{a}(\theta)^H\mathbf{L}\mathbf{L}^H\mathbf{a}(\theta)} = \frac{1}{|\mathbf{a}(\theta)^H\mathbf{L}|^2} = \frac{1}{\sum_{k=1}^N|\mathbf{a}(\theta)^H l_{*k}|^2}, \tag{74}$$

$$P_{ME}(\theta) = \frac{\mathbf{1}_1^T\mathbf{L}\mathbf{L}^H\mathbf{1}_1}{|\mathbf{a}(\theta)^H\mathbf{L}\mathbf{L}^H\mathbf{1}_1|^2} = \frac{|l_{11}|^2}{|\mathbf{a}(\theta)^H \boldsymbol{l}_{*1}l_{11}|^2} = \frac{1}{|\mathbf{a}(\theta)^H\boldsymbol{l}_{*1}|^2}, \tag{75}$$

where $\boldsymbol{l}_{*k}$ denotes the $k$-th column of the matrix $\mathbf{L}$.

A comparison of (74) and (75) reveals the relationship between the Capon and maximum entropy estimates:

$$P_{CP}(\theta) = \frac{1}{\sum_{k=1}^N \frac{1}{P_{ME,k}(\theta)}}, \tag{76}$$

where $P_{ME,k}(\theta)$ is the autoregressive estimate of the maximum entropy spectrum of order k.

The estimate $P_{ME,k}(\theta)$ corresponds to the standard maximum entropy spectrum of a subarray formed by the first $k$ elements of the original $N$-element antenna:

$$P_{ME,k}(\theta) = \frac{\mathbf{1}_1^T\mathbf{R}_k^{-1}\mathbf{1}_1}{|\mathbf{a}(\theta)^H\mathbf{R}_k^{-1}\mathbf{1}_1|^2}, \tag{77}$$

where $\mathbf{R}_k^{-1}$ is the inverse covariance matrix of the first $k$ elements subarray.

Therefore, the Capon spectrum estimate represents the harmonic mean of the maximum entropy spectral estimates across all model orders from 1 to $N$. This explains the lower resolution of the Capon method. While the denominator in the maximum entropy spatial spectrum (75) corresponds to the adaptive pattern of a full $N$-element antenna array, the denominator in the Capon spatial spectrum (74) represents the sum of adaptive patterns across arrays of sizes 1, 2, …, $N$. The lower-order patterns smooth the aggregate adaptive pattern, broadening the main beam of the Capon spectrum.

Fig. 6 compares the maximum entropy and Capon estimates, interpreting both as methods for extrapolating the correlation function beyond the observed window. Here, $F_\infty$ denotes the Fourier transform of an infinite sequence. A $k$-order recursive filter represents a linear feedback filter that generates a $k$-order autoregressive process. The filter coefficients are defined as the elements in the first column of the inverse covariance matrix of the corresponding order, while the initialization values are the corresponding correlation coefficients.

Generating an infinite correlation sequence and computing its spatial spectrum via the Fourier transform is mathematically equivalent to convolving the autoregressive parameters with steering vectors across all directions in according to (74) and (75). Fig. 6 illustrates the conceptual interpretation of the maximum entropy and Capon methods regarding correlation function extrapolation, rather than their specific computational steps.

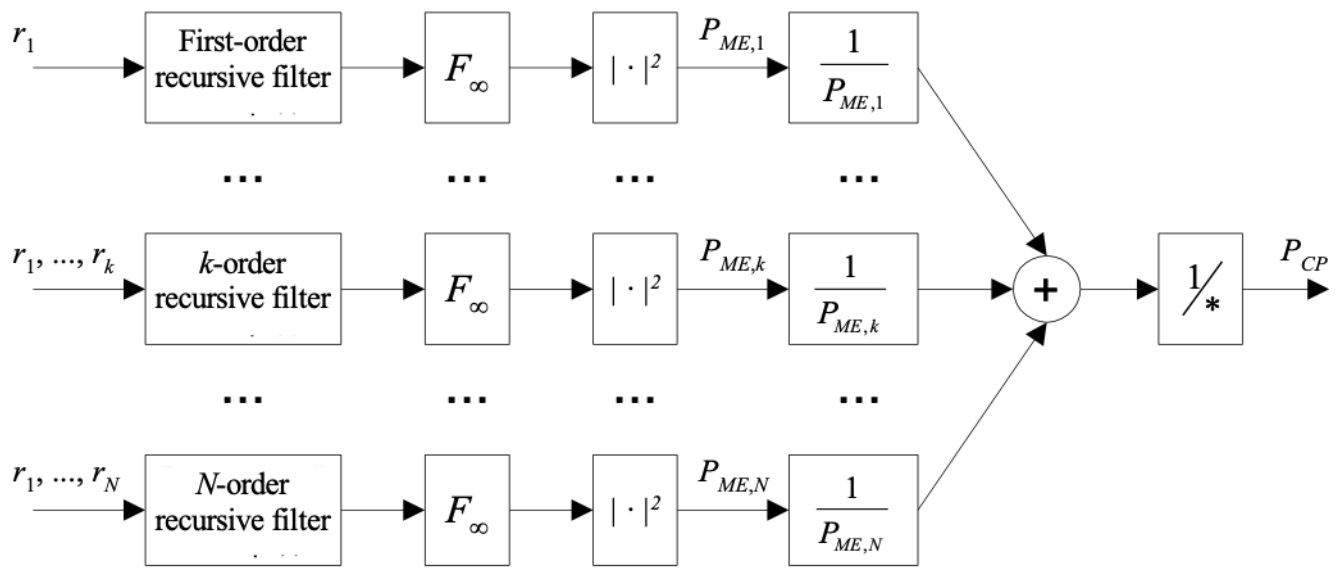

Fig. 6. Connection between maximum entropy and Capon spectrum estimates.

Fig. 7 illustrates the relationship between the maximum entropy and Capon spectra defined in (76). The Capon spectrum in Fig. 7(i) is obtained by averaging the maximum entropy spectra across antenna subarrays with 1 to 16 elements.

The graphs in Fig. 7 illustrate the relationship between the maximum entropy and Capon spectra, as defined in (76), showing that the Capon spectrum presented in Fig. 7(i) is derived by averaging maximum entropy spectra across antenna subarrays with elements ranging from 1 to 16. While this averaging process reduces sensitivity to antenna array calibration errors, it degrades spatial resolution, failing to resolve sources at coordinates 0.15 and 0.22.

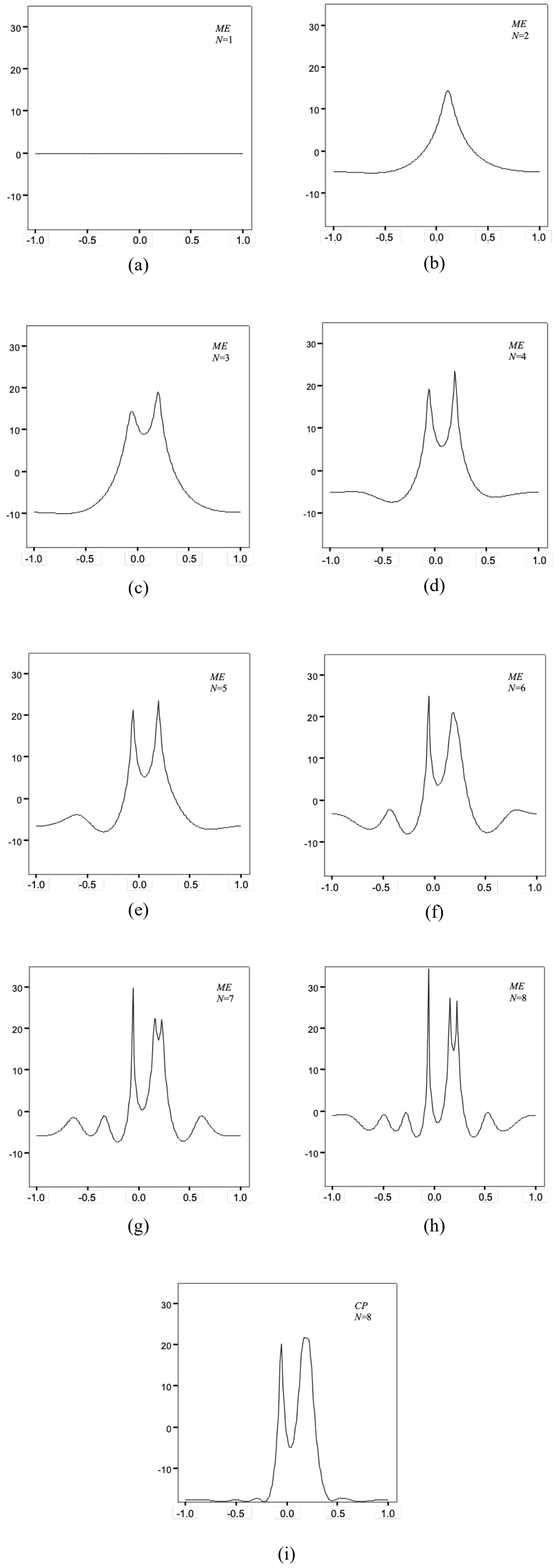

Fig. 7. Maximum entropy spatial spectra for antenna arrays of 1 to 8 sensors (a–h) and the Capon spectrum for an 8-element array (i). Three equal-power signal sources are located at coordinates -0.06, 0.15, and 0.22.

Next, we consider MUSIC method, a direct generalization of the Pisarenko approach. Their respective spatial spectrum estimates are defined as:

$$P_{MUSIC}(\theta) = \frac{1}{\mathbf{a}(\theta)^H \mathbf{E}_{N-d}\mathbf{E}_{N-d}^H \mathbf{a}(\theta)}, \tag{78}$$

$$P_{PSR}(\theta) = \frac{1}{|\mathbf{a}(\theta)^H \mathbf{e}_1|^2}, \tag{79}$$

where $\mathbf{E}_{N-d} = [\mathbf{e}_1, \mathbf{e}_2, \ldots, \mathbf{e}_{N-d}]$ is an $N \times (N-d)$ matrix whose columns are the noise eigenvectors corresponding to the repeated noise eigenvalue $\sigma^2$ of the covariance matrix **R**.

The MUSIC spectrum can be also expressed as

$$P_{MUSIC}(\theta) = \frac{1}{\sum_{k=1}^{N-d} |\mathbf{a}(\theta)^H \mathbf{e}_k|^2}. \tag{80}$$

While both MUSIC and Pisarenko are subspace methods, viewing MUSIC as an extrapolation of the autocorrelation function requires distinct analyses for single repeated noise eigenvalues and unique noise eigenvalues.

First, we consider the case where the covariance matrix has a Toeplitz structure, and its spectrum contains a single repeated noise eigenvalue. We show that, under these conditions, the Pisarenko and MUSIC methods yield equivalent estimates.

The noise subspace projection matrix $\mathbf{E}_{N-d}\mathbf{E}_{N-d}^H$ of the covariance matrix in (85) can be represented as:

$$\mathbf{E}_{N-d}\mathbf{E}_{N-d}^H = \mathbf{E}_{N-d}\mathbf{Q}^H \mathbf{Q}\mathbf{E}_{N-d}^H = \mathbf{W}\mathbf{W}^H, \tag{81}$$

where $\mathbf{Q}$ be an $(N\text{-}d)\times(N\text{-}d)$ unitary matrix ($\mathbf{Q}^H\mathbf{Q} = \mathbf{I}$) that transforms the noise eigenvectors of the covariance matrix into a lower triangular matrix **W** presented in Fig. 8. Although the columns of **W** are not mutually orthogonal, they still span the noise subspace of the covariance matrix.

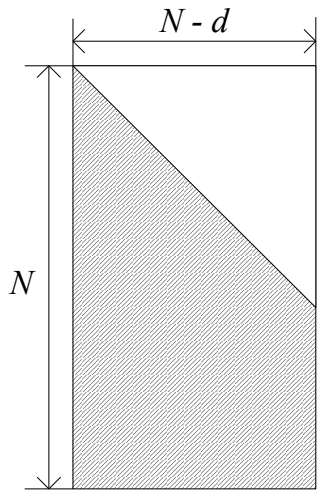

Fig. 8. Matrix **W** of the transformed noise eigenvectors.

Since a unitary transformation of the eigenvectors preserves the underlying subspace, the MUSIC spectrum can be expressed as:

$$P_{MUSIC}(\theta) = \frac{1}{\mathbf{a}(\theta)^H \mathbf{W}\mathbf{W}^H \mathbf{a}(\theta)} = \frac{1}{\sum_{k=1}^{N-d} |\mathbf{a}(\theta)^H w_k|^2}, \tag{82}$$

where $\mathbf{w}_k$ is the $k$-th column of the matrix **W**.

Since **W** is triangular, the subvector of non-zero elements in column $\mathbf{w}_k$, (omitting the first) serves as an eigenvector for the subarray of the last ($N$-$k$+1) antenna elements. This relationship is written as:

$$\mathbf{R}^{(k)}\mathbf{W}^{(k)} = \sigma^2 \mathbf{W}^{(k)}, \tag{83}$$

where $\mathbf{R}^{(k)}$ and $\mathbf{W}^{(k)}$ are the lower-right submatrices of **R** and **W**, with dimensions $(N\text{-}k+1)\times(N\text{-}k+1)$ and $(N\text{-}k+1)\times(N\text{-}d\text{-}k+1)$, respectively,

$$\mathbf{W}^{(k)} = \begin{bmatrix} \mathbf{0}_{k-1} & \mathbf{0} \\ \mathbf{0} & \mathbf{I}_{N-k+1} \end{bmatrix} \mathbf{W} \begin{bmatrix} \mathbf{0}_{k-1} & \mathbf{0} \\ \mathbf{0} & \mathbf{I}_{N-d-k+1} \end{bmatrix}, \tag{84}$$

$$\mathbf{R}^{(k)} = \begin{bmatrix} \mathbf{0}_{k-1} & \mathbf{0} \\ \mathbf{0} & \mathbf{I}_{N-k+1} \end{bmatrix} \mathbf{R} \begin{bmatrix} \mathbf{0}_{k-1} & \mathbf{0} \\ \mathbf{0} & \mathbf{I}_{N-k+1} \end{bmatrix}. \tag{85}$$

As established above, the noise eigenvector of the Toeplitz covariance matrix defines an autoregressive extrapolation of the correlation function beyond the antenna aperture. Consequently, each noise eigenvector $\mathbf{w}_k$ of dimension ($N$-$k$+1) defines a minimum-entropy extrapolation for the ($N$-$k$+1)-element subarray. The autoregressive equation describing this extension is given by:

$$r_{s+1}^{(k)} = -\frac{1}{w_{N-k+1}^{(k)}} \sum_{m=1}^{N-k} w_m^{(k)} r_{s+i-N+k}^{(k)}, \tag{86}$$

where $r_{s+1}^{(k)}$ is the ($s$+1)-th element in the continued correlation function of the $k$-th subarray, which consists of $N$-$k$+1 elements, and $\boldsymbol{w}^{(k)} = \left[w_1^{(k)}, w_2^{(k)}, \ldots, w_{N-k+1}^{(k)}\right]^T$ denotes the noise eigenvector of the $k$ -th subarray.

We now derive the relationship between the autoregressive coefficients $w_i^{(k)}$ in (86) for the Pisarenko and MUSIC methods under the assumption of a repeated noise eigenvalue.

The initial values of the correlation sequences in (86) are known and identical. Because the diagonal elements are not used in the prediction, these values can be drawn from either the covariance matrix **R** or its noise-free counterpart $\mathbf{C} = \mathbf{R} - \sigma^2\mathbf{I}.$

According to the singular extension theorem for Toeplitz covariance matrices [11], an $N\times N$ singular Toeplitz matrix **C** of rank $\underline{d} < N$ has a unique singular extension, provided its principal minor of order $d$ is non-zero. The recurrence formula (86) specifies $(N-d)$ autoregressive extensions of the correlation function and their corresponding singular extensions. To verify the singularity of these extensions, we evaluate one step of the recurrence process for the autoregressive coefficient vector $\mathbf{w}_k$. By assumption, $\mathbf{w}_k$ is the eigenvector corresponding to the zero eigenvalue of the matrix $\mathbf{C}^{(k)} = \mathbf{R}^{(k)} - \sigma^2\mathbf{I}$:

$$\begin{bmatrix} c_1 & c_2 & \cdots & c_{N-k+1} \\ c_2^* & c_1 & \cdots & c_{N-k} \\ \cdots & \cdots & \cdots & \cdots \\ c_{N-k+1}^* & c_{N-k}^* & \cdots & c_1 \end{bmatrix} \begin{bmatrix} w_1 \\ w_2 \\ \cdots \\ w_{N-k+1} \end{bmatrix} = \begin{bmatrix} 0 \\ 0 \\ \cdots \\ 0 \end{bmatrix}. \tag{87}$$

The extended matrix will be singular if its next extension—$(N-k+2)$-th column—is a linear combination of the preceding ones:

$$\begin{bmatrix} c_2 & \cdots & c_{N-k+1} & c_{N-k+2} \\ c_1 & \cdots & c_{N-k} & c_{N-k+1} \\ \cdots & \cdots & \cdots & \cdots \\ c_{N-k+1}^* & \cdots & c_1 & c_2 \end{bmatrix} \begin{bmatrix} w_1 \\ w_2 \\ \cdots \\ w_{N-k+1} \end{bmatrix} = \begin{bmatrix} 0 \\ 0 \\ \cdots \\ 0 \end{bmatrix}. \quad (88)$$

The orthogonality of vector $\mathbf{w}_k$ to rows 2 through $N-k+1$ of the matrix in (88) follows directly from (87). Its orthogonality to the first row is a consequence of the recurrence formula (86) for the predicted matrix element. Consequently, each eigenvector (the vector of autoregression coefficients) $\mathbf{w}_k$ determines a singular extension of the noise-free covariance matrix for the $(N-k+1)$-element subarray.

While there exist $N-d$ infinite singular Toeplitz matrices of rank $d$ that share a $(d+1)\times(d+1)$ block $\mathbf{C}^{(d+1)}$, they represent different extensions of the same submatrix $\mathbf{C}^{(d+1)}$. The singular extension theorem, however, guarantees uniqueness of the extension under these conditions. Therefore, all resulting autoregressive correlation sequences and their spectra are identical.

Let us now formulate the relationship between the Pisarenko and MUSIC spatial spectra. We define the spectrum of the $k$-th correlation sequence, which is generated by the noise eigenvector $\mathbf{w}_k$ in accordance with (93), as

$$P_{PSR,k}(\theta) = \frac{1}{|\mathbf{a}(\theta)^H \mathbf{w}_k|^2}. \quad (89)$$

Then expression (89) defining the MUSIC spatial spectrum estimate takes the following form:

$$P_{MUSIC}(\theta) = \frac{1}{\sum_{k=1}^{N-d} \frac{1}{P_{PSR,k}(\theta)}}. \quad (90)$$

Thus, the MUSIC spectrum can be viewed as the harmonic mean of ($N$-$d$) Pisarenko spectra, while the Capon spectrum similarly equates to the harmonic mean of the maximum entropy estimates.

Since the subarray Pisarenko spectra of a uniform antenna array are identical, their average in the MUSIC estimate offers no advantage over a single Pisarenko spectrum. Thus, MUSIC and the Pisarenko method are identical when evaluating a single multiple noise eigenvalue. In the presence of a single multiple noise eigenvalue, the MUSIC spectral estimate becomes mathematically equivalent to the Pisarenko estimate.

While the MUSIC and Pisarenko spectra generally differ when noise eigenvalues vary or the antenna array is non-equidistant, the relationship in (90) holds true regardless of array geometry. However, for non-equidistant arrays, these spectra can no longer be interpreted as minimum entropy extensions of the correlation function.

The relationship between the MUSIC spectrum estimates and the minimum-entropy prediction of the correlation function is schematically illustrated in Fig. 9.

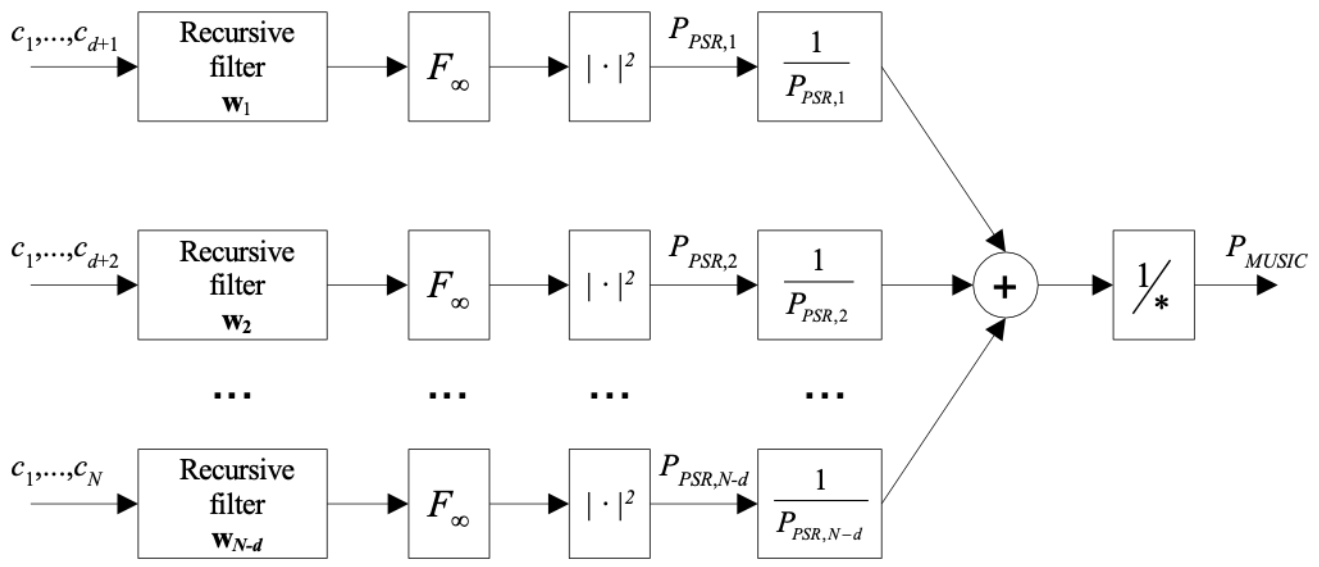

Fig. 9. Relationship between Pisarenko and MUSIC spectrum estimates.

Fig. 9 compares the Pisarenko and MUSIC estimates from a conceptual standpoint, rather than focusing on specific computational scheme. To produce the Pisarenko spectral estimates, an equivalent inverse (non-recursive) filter with matching coefficients is used. Fig. 10 illustrates two equivalent definitions of the Pisarenko spectrum.

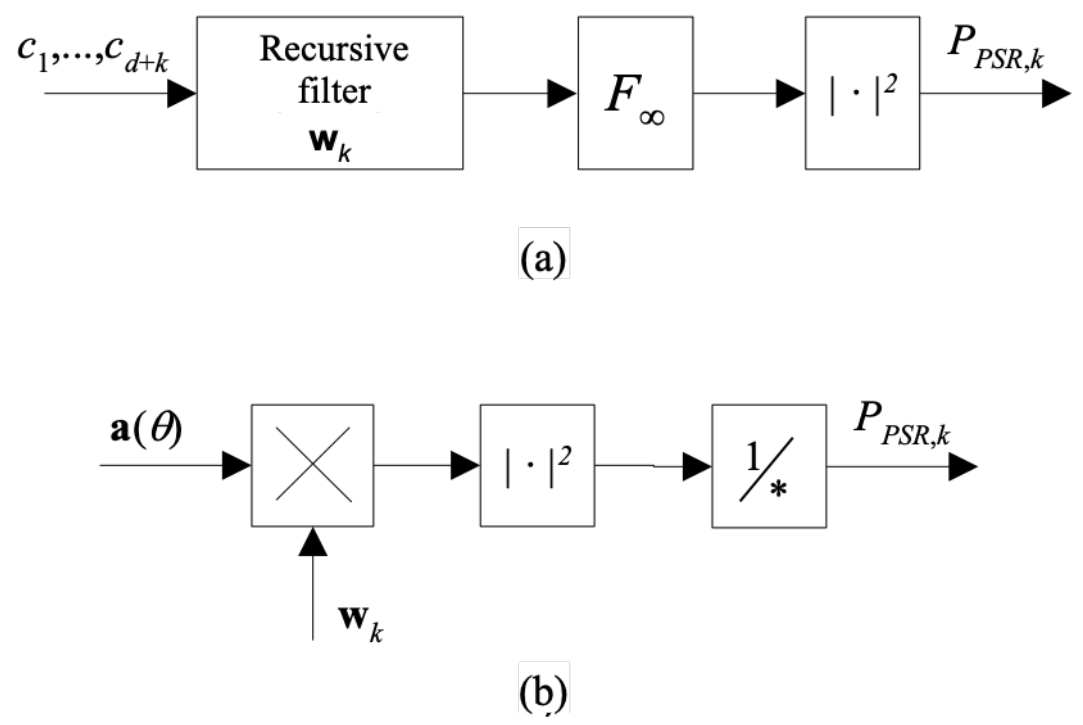

Fig. 10. Two definitions of the Pisarenko spectrum: (a) the Fourier transform of the infinite correlation sequence; (b) the inverse squared magnitude of the convolution of the noise eigenvector $\mathbf{w}_k$ and steering vectors $\mathbf{a}(\theta)$

Thus, the Capon and MUSIC methods represent harmonically averaged estimates of the spectra of infinite extensions of the correlation function beyond the observed region, corresponding to maximum entropy and minimum entropy (Pisarenko) extensions, respectively. For MUSIC and Pisarenko, these extensions equate to predicting the plane wavefronts of sources beyond the physical antenna aperture.

## VI. A Unified Framework for Super-Resolution Methods

### A. Signal Models Spanning the Antenna Array and Beyond

In previous sections, we interpreted angular super-resolution methods as extrapolating received signal correlation functions or source wavefronts beyond the physical antenna aperture. To facilitate a comparison, Fig. 11 provides a schematic summary of this concept. It illustrates signal models for the traditional and super-resolution methods, depicting linear antenna arrays and wavefronts both within and beyond the aperture for a single signal source.

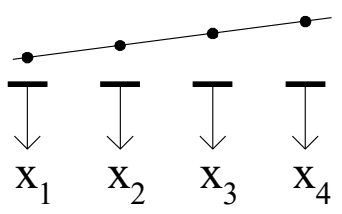

(a) Classical methods (no signal assumptions beyond the antenna aperture)

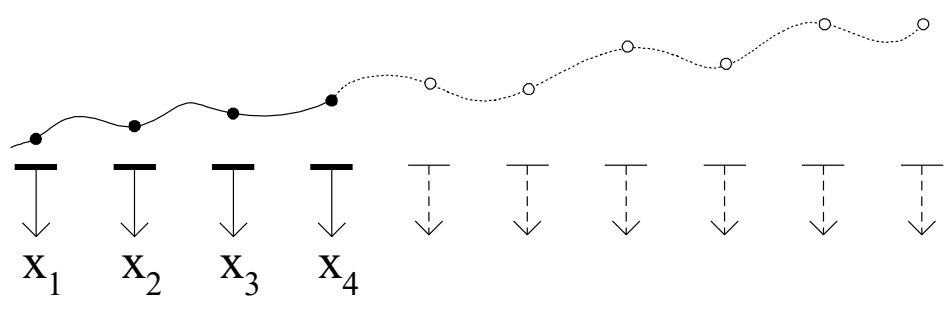

(b) Maximum entropy

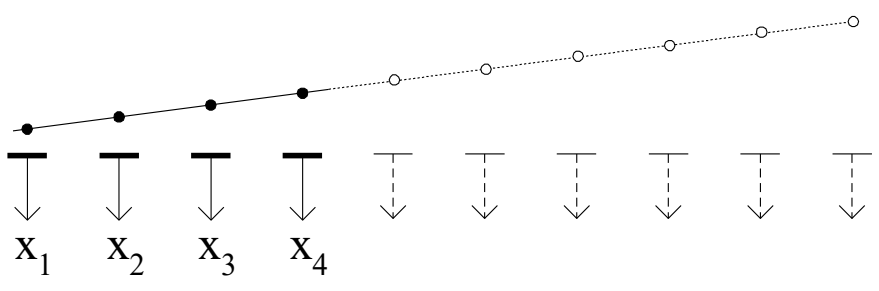

(c) Pisarenko method: Minimum entropy

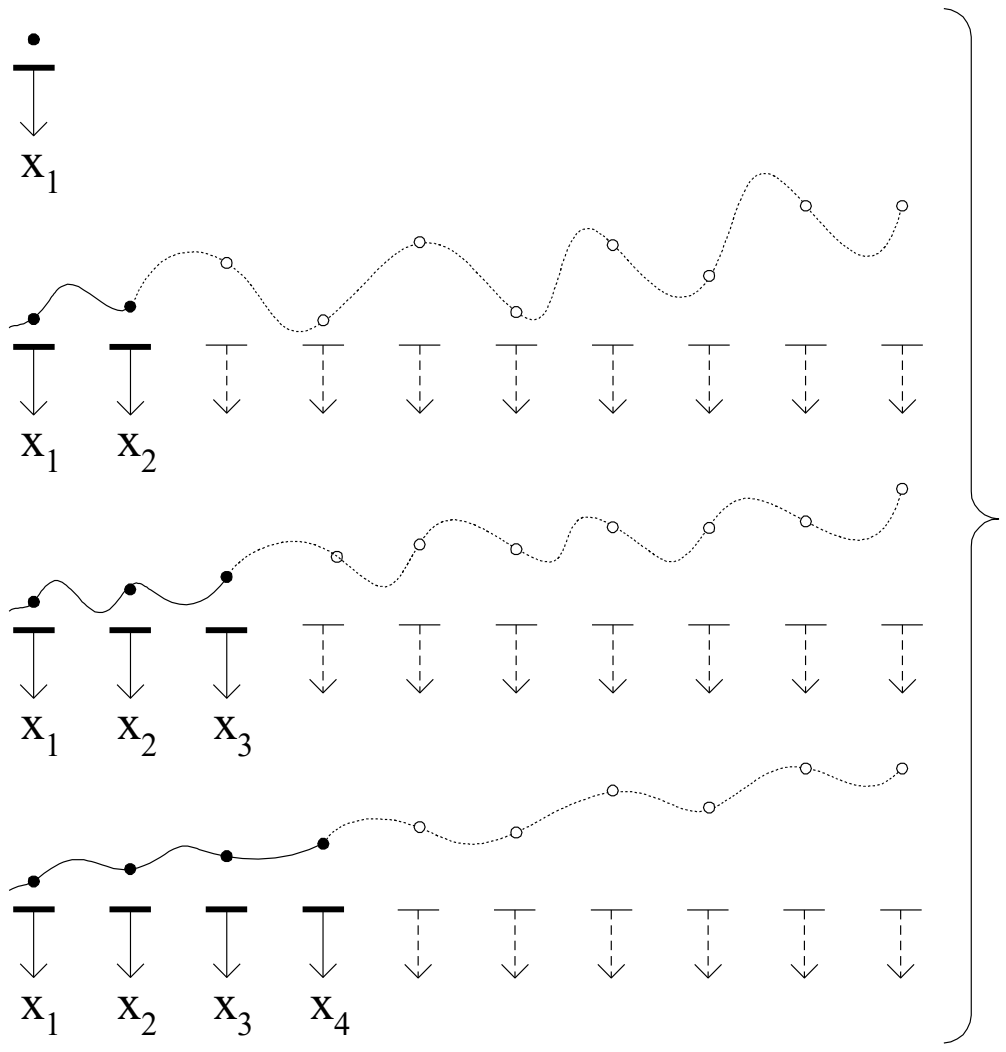

(d) Capon: Averaging the maximum entropy estimates for all subarray sizes from 1 to *N*

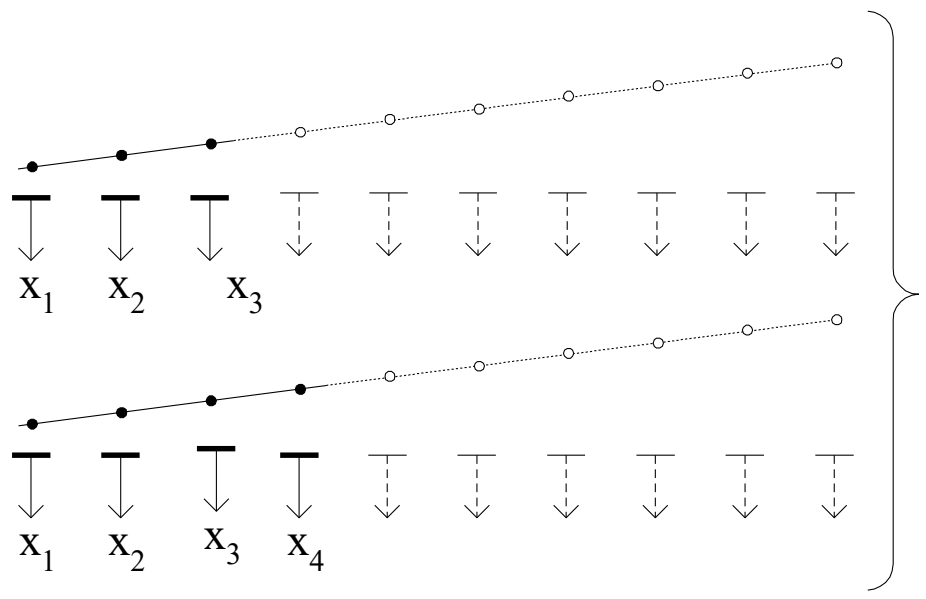

(e) MUSIC: Averaging of the minimum entropy (Pisarenko) estimates

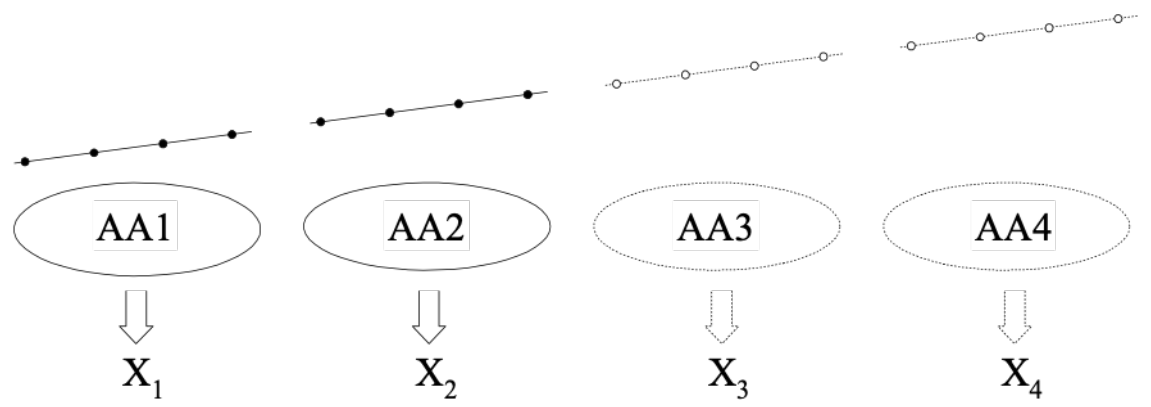

(f) ESPRIT: Linear prediction by subarrays using minimum entropy criterion

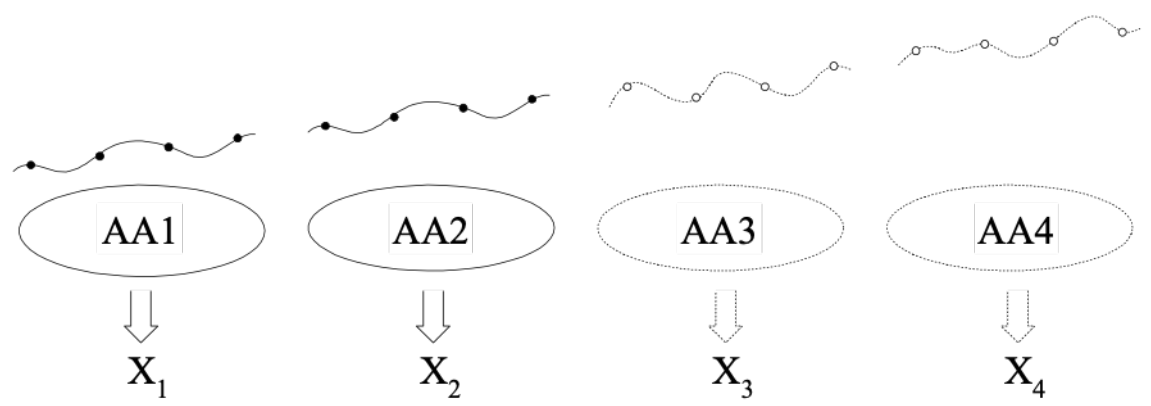

(g) Approximate ESPRIT: Linear prediction by subarrays using maximum entropy criterion

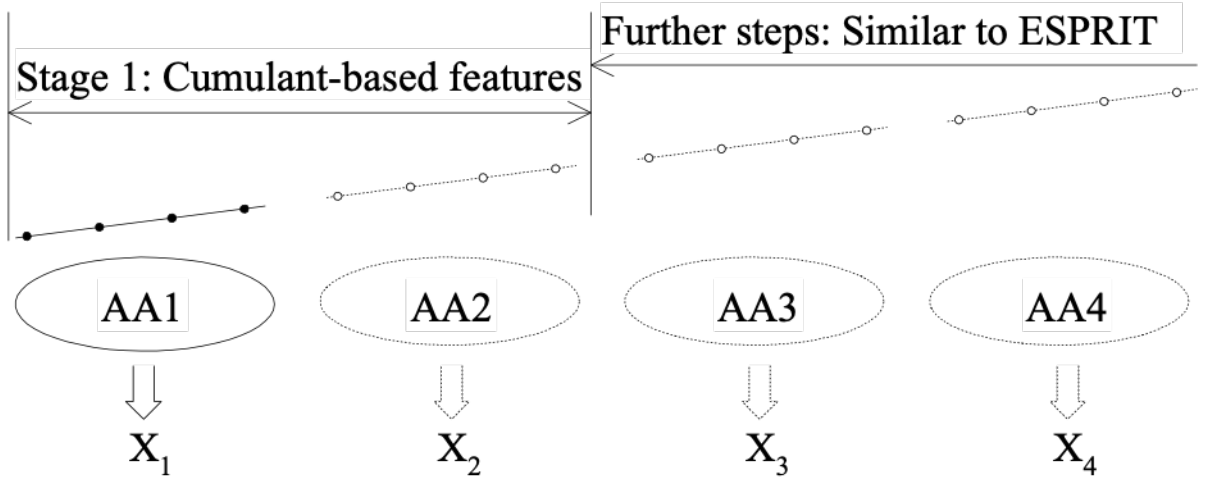

(h) VESPA: Linear prediction by subarrays using the "minimum entropy" criterion with higher-order statistics

Fig. 11. Signal models at and beyond the antenna aperture.

Classical angular measurement methods utilize a plane-wave model for signal sources within the antenna aperture and zero elsewhere as presented in Fig. 11(a). Variations of this approach incorporate spatial smoothing windows, which taper the signal amplitude transition at the aperture edges. This technique suppresses false peaks and sidelobe levels, reducing angular resolution.

Super-resolution algorithms that surpass the Rayleigh limit differ primarily in how they model wavefronts beyond the antenna aperture. Regardless of the specific technique, the ultimate performance of any super-resolution method depends on two key factors: how it extrapolates the correlation function outside the aperture, and how closely its underlying assumptions match the physical environment.

The maximum entropy method, illustrated in Fig. 11(b), extrapolates the source wavefronts beyond the antenna aperture maximizing the entropy of their correlation function. Conversely, the Pisarenko method minimizes this entropy, thereby extending plane wavefronts beyond the aperture as shown in Fig. 11(c).

Fig. 11(d) shows the Capon signal model that averages *N* maximum entropy estimates of the source wavefronts, using antenna subarrays ranging from 1 to *N* elements.

Fig. 11(e) shows the interpretation of the MUSIC method as extending plane signal wavefronts beyond the physical array. In the presence of d signal sources, N-d correlation functions are extrapolated using the minimum entropy criterion. Each extrapolated function corresponds to extending the plane wavefronts of all d signal sources. The resulting MUSIC spatial spectrum is the harmonic average of the spectra derived from the extended correlation functions.

Operating on two identical subarrays, the ESPRIT method can be interpreted as extrapolating the plane wavefronts of signal sources beyond the physical array, as shown in Fig. 11(f). This wavefront extrapolation is equivalent to applying a minimum entropy criterion to the extended blocks of the correlation function. The subarrays are equidistantly spaced. For conceptual clarity, they are depicted as non-overlapping, though they may overlap.

Fig. 11(g) illustrates a maximum entropy variant of ESPRIT, which we call Approximate ESPRIT. This variant retains the classical ESPRIT array configuration but utilizes a maximum entropy extension of the correlation function.

Fig. 11(h) illustrates the linear prediction interpretation of the VESPA algorithm. In the first stage, fourth-order cumulants are used to synthesize a virtual secondary antenna array. The second stage then extrapolates the plain signal wavefronts to subsequent predicted virtual subarrays, mirroring the standard ESPRIT approach.

### *B. Common Space for Angular Super-resolution Methods*

We formulate angular super-resolution methods using a unified framework based on linear prediction and correlation-function extrapolation beyond the observed domain. This approach defines a multidimensional space of super-resolution estimates, whose axes represent entropy, averaging of estimates, and multichannel subarray prediction.

The origin of the estimation space (Fig. 12) represents the Pisarenko method. This point features a minimum-entropy extrapolated correlation function, single-channel prediction, and no averaging of the estimates.

The extreme opposite of the entropy axis represents the maximum entropy method for single-channel, unaveraged prediction. Estimates with intermediate entropy values lie between the Pisarenko and maximum entropy limits.

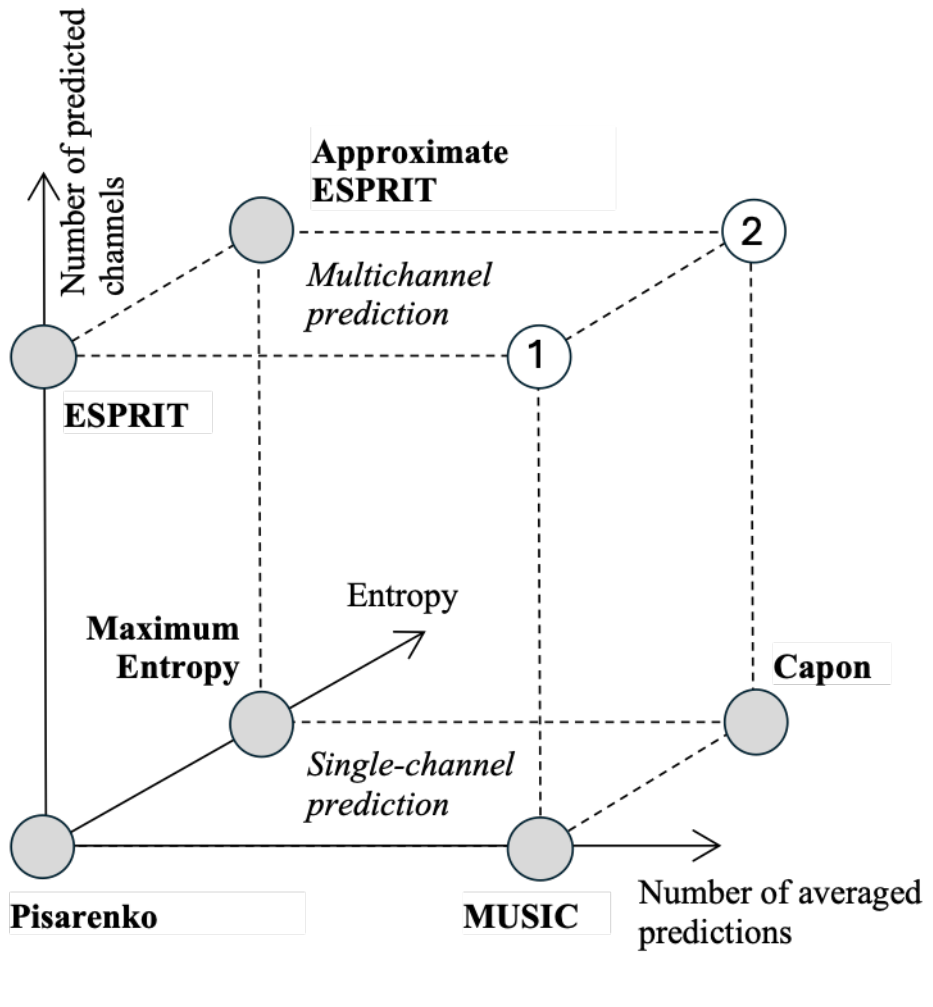

Fig 12. Common space for angular super-resolution methods.

The harmonic averaging of correlation function predictions corresponds to the MUSIC estimate for minimum-entropy predictions and the Capon estimate for maximum-entropy predictions. Specifically, the number of averaged predictions equals the number of signal sources $(N-d)$ for the MUSIC method, and the number of antenna array channels ($N$) for the Capon method. However, our graphical analysis indicates the presence or absence of averaging, without differentiating between the specific positions of these methods along the "Number of averaged predictions" axis.

The upper face of the cube represents the space of multichannel prediction by subarrays. The boundary methods in this plane are multidimensional counterparts to traditional single-channel techniques; for example, the ESPRIT algorithm corresponds to Pisarenko minimum entropy method, while approximate ESPRIT parallels the maximum entropy method. Points 1 and 2 designate the boundary states that have not yet been addressed by existing super-resolution algorithms. Details on deriving estimate 1 are provided below.

Our representation of the Pisarenko, maximum entropy, MUSIC, and Capon methods in the single-channel forecast plane is conceptually similar to the approach proposed in [27], which utilizes the signal-to-noise ratio rather than the entropy of the predicted correlation function. More importantly, however, the proposed angular super-resolution space not only establishes the relationships among these key method groups but also enables smooth transitions between them. This allows for the generation of arbitrary intermediate estimates that lie between established techniques.

Any point within the cubic region can be assigned an estimate, making this domain the estimate space for angular super-resolution methods. In Section VII we show how to practically obtain intermediate estimates bounded by this space. Furthermore, we introduce specific algorithms that enable for seamless transitions between estimates within a unified computational framework.

We now extend our super-resolution estimate space to higher-order statistics, requiring a redefinition of entropy. Since the standard log-determinant of a covariance matrix is valid only for Gaussian processes, evaluating non-Gaussian signals requires returning to the fundamental probability density definition. Without specifying this density, entropy cannot be directly expressed in terms of cumulant matrices. However, for extended signals, we are not interested in the exact entropy, but rather in how it reflects the nature of the spatial spectrum. Therefore, we introduce a surrogate measure analogous to the Gaussian covariance matrix determinant.

The cumulant matrix, used in the VESPA method to estimate the angular coordinates of sources, is the sum of the normalized source covariance matrices multiplied by specific scaling factors. In other words, the VESPA method leverages fourth-order statistics to calculate second-order statistics, which are then used to estimate the spectra or angular coordinates of the signal sources. In the presence of non-Gaussian uncorrelated noise, this cumulant matrix takes the form:

$$\widetilde{\mathbf{R}} = \sum_{i=1}^{d} \beta_i \mathbf{R}_i = \mathbf{A\Gamma A}^H + \sigma^2 \mathbf{I}, \tag{91}$$

where $\mathbf{R}_i$ is the covariance matrix of the $i$-th source signal across the antenna aperture; $\beta_i = \gamma_i / \sigma_i^2$ is a scaling factor defined as the ratio of the fourth-order cumulant to the signal variance; $\gamma_i$ and $\sigma_i^2$ are the fourth-order cumulant and variance of the $i$-th source signal, respectively; $\mathbf{\Gamma} = \mathrm{diag}\{\gamma_i\}$ is the diagonal matrix of these cumulants; and $\sigma^2$ is the variance of the uncorrelated non-Gaussian noise in the receiving channels.

We use the determinant of the cumulant matrix $\widetilde{\mathbf{R}}$ as the third dimension in our space of angular super-resolution estimates based on fourth-order statistics.

As discussed in Section IV, the VESPA method performs multichannel prediction utilizing the minimum cumulant matrix determinant criterion without averaging of the estimates. Fig. 13 illustrates the overall space of fourth-order super-resolution estimates, within which the VESPA method represents the only filled boundary cell. The remaining vertices bounding the cumulant estimate space are defined by the natural physical limits of the key parameters: the full cumulant matrix determinant, the number of simultaneously predicted channels, and the number of averaged predictions.

In the single-channel prediction plane, Fig. 13 illustrates boundary estimates for non-Gaussian signals, mirroring the second-order statistics approach shown in Fig. 12. Clearly, using a cumulant matrix instead of a covariance matrix allows for the development of higher-order counterparts to standard correlation-based methods.

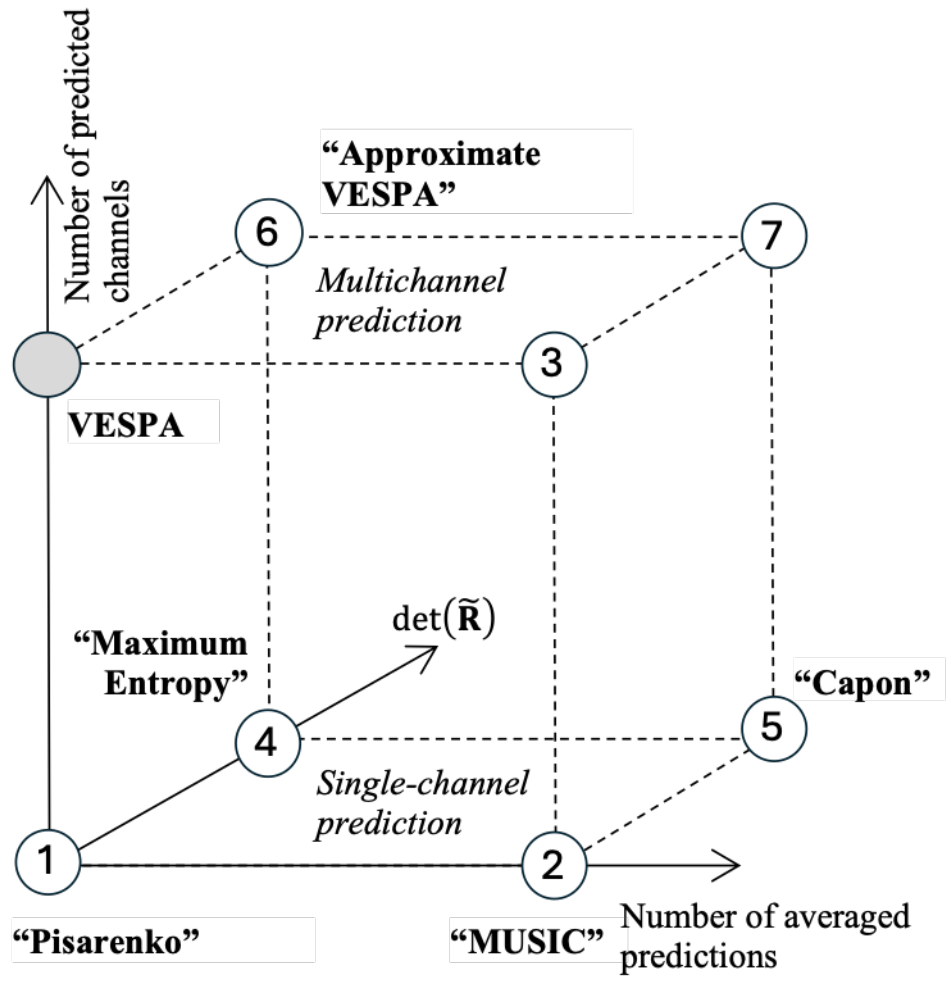

Fig. 13. Common space for angular super-resolution methods of non-Gaussian signals.

### C. An Example of Synthesizing the ‘Missing’ Super-resolution Method

Some cube vertices in the angular super-resolution space (Fig. 12) are empty. To show that a corresponding super-resolution method can be derived for every vertex or internal point, we get an estimate for the method labeled 1.

Spatial spectrum estimates for unfilled cube vertices can be synthesized via adjacent edge methods. For vertex 1, you can use MUSIC (with a multichannel version) or ESPRIT (with harmonic averaging). We focus on the ESPRIT approach.

Standard ESPRIT uses only two identical subarrays. Applying harmonic averaging, however, requires more than two subarrays. We therefore use $K$ identical physical subarrays, labeled as $\mathrm{AP}_1 \ldots \mathrm{AP}_K$ in Fig. 14 (which illustrates the signal models on and beyond the antenna aperture).

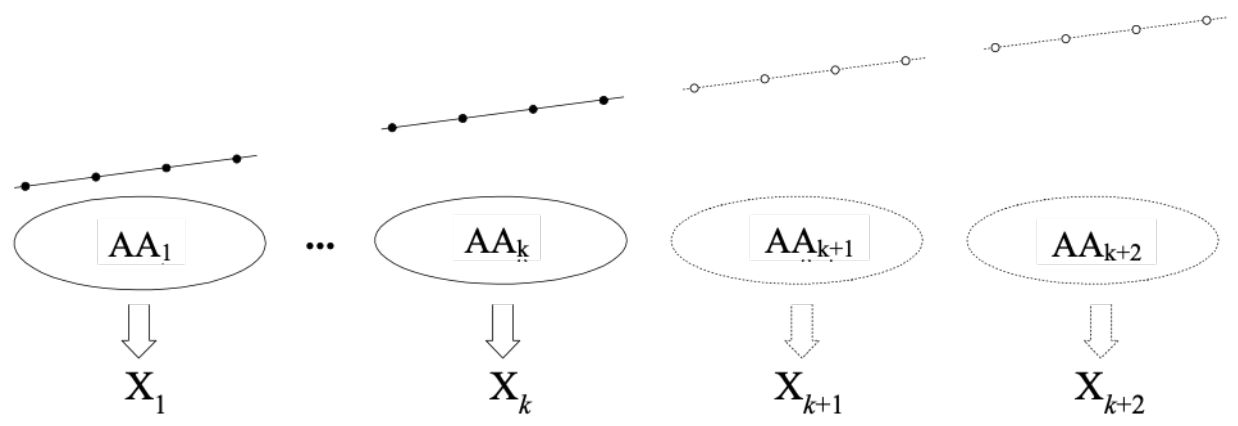

Fig. 14. Minimum entropy linear prediction using subarrays and harmonic averaging.

Without source cross-correlation, the real antenna array's covariance matrix for ESPRIT becomes:

$$\mathbf{C} = \begin{bmatrix} \mathbf{C}_1 & \mathbf{C}_2 & \cdots & \mathbf{C}_k \\ \mathbf{C}_1^H & \mathbf{C}_1 & \cdots & \mathbf{C}_{k-1} \\ \cdots & \cdots & \cdots & \cdots \\ \mathbf{C}_k^H & \mathbf{C}_{k-1}^H & \cdots & \mathbf{C}_1 \end{bmatrix}, \tag{92}$$

where $\mathbf{C}_i$ is the mutual covariance matrix of signals of the $m$-th and $(m+i-1)$-th subarrays; $K$ is the number of real subarrays.

We assume the number of signal sources $d$ satisfies $d \leq N$, where $N$ is the dimension of matrix $\mathbf{C}_i$ (the number of channels in a subarray). Following Section III, the generalized eigenvalues of matrix pencils formed by adjacent matrix pairs $\mathbf{C}_i$ and $\mathbf{C}_{i+1}$ matrix $\mathbf{C}$ are identical. These eigenvalues directly yield the angular directions of the signal sources:

$$\mathbf{C}_1 \mathbf{U} = \mathbf{C}_{i+1} \mathbf{U} \mathbf{\Lambda}, \tag{93}$$

where $\mathbf{U} = [\mathbf{u}_1, \mathbf{u}_2, \ldots, \mathbf{u}_N]$, and $\mathbf{\Lambda} = \mathrm{diag}\{\lambda_m\}_{m=1,\ldots,N}$, $\mathbf{u}_m$, $\lambda_m$ represents the $m$-th generalized eigenvector and the $m$-th generalized eigenvalue of the beam, respectively. The angular coordinates of the sources relate to the generalized eigenvalues as follows:

$$\mathbf{\Lambda} = \mathrm{diag}\{\exp(j2\pi|\Delta|\sin\theta_1), \ldots, \exp(j2\pi|\Delta|\sin\theta_d), 0, \ldots, 0\}. \tag{94}$$

When matrices $\mathbf{U}$ and $\mathbf{\Lambda}$ satisfy equation (93), they can be used to construct a matrix of eigenvectors for $\mathbf{C}$ corresponding to the zero eigenvalue. The block columns of this matrix, each containing $K$ blocks of size $N \times N$, can then be interpreted as the block eigenvectors of the noise-free block covariance matrix $\mathbf{C}$:

$$\mathbf{CB} = \mathbf{0} \tag{95}$$

or

$$\begin{bmatrix} \mathbf{C}_1 & \mathbf{C}_2 & \mathbf{C}_3 & \dots & \mathbf{C}_{K-1} & \mathbf{C}_K \\ \mathbf{C}_2^H & \mathbf{C}_1 & \mathbf{C}_2 & \dots & \mathbf{C}_{K-2} & \mathbf{C}_{K-1} \\ \mathbf{C}_3^H & \mathbf{C}_2^H & \mathbf{C}_1 & \dots & \mathbf{C}_{K-3} & \mathbf{C}_{K-2} \\ \dots & \dots & \dots & \dots & \dots & \dots \\ \mathbf{C}_{K-1}^H & \mathbf{C}_{K-2}^H & \mathbf{C}_{K-3}^H & \dots & \mathbf{C}_1 & \mathbf{C}_2 \\ \mathbf{C}_K^H & \mathbf{C}_{K-1}^H & \mathbf{C}_{K-2}^H & \dots & \mathbf{C}_2^H & \mathbf{C}_1 \end{bmatrix} \begin{bmatrix} \mathbf{B}_1 & \mathbf{0} & \dots & \mathbf{0} \\ \mathbf{B}_2 & \mathbf{B}_1 & \dots & \mathbf{0} \\ \mathbf{0} & \mathbf{B}_2 & \dots & \mathbf{0} \\ \dots & \dots & \dots & \dots \\ \mathbf{0} & \mathbf{0} & \dots & \mathbf{B}_1 \\ \mathbf{0} & \mathbf{0} & \dots & \mathbf{B}_2 \end{bmatrix} = \begin{bmatrix} \mathbf{0} \\ \mathbf{0} \\ \mathbf{0} \\ \dots \\ \mathbf{0} \\ \mathbf{0} \end{bmatrix},$$

where $\mathbf{B}_1$ and $\mathbf{B}_2$ are $N \times N$ square matrices. Since these matrices are not uniquely determined, one valid choice satisfying equation (93) is:

$$\mathbf{B}_1 = \mathbf{U}; \quad \mathbf{B}_2 = -\mathbf{U}\boldsymbol{\Lambda}. \tag{96}$$

Each block eigenvector forms a multichannel prediction vector. When the full antenna array has a block-Toeplitz covariance matrix, these predictions match. If cross-correlated sources break this block-Toeplitz structure, the predictions diverge, changing the structure of matrix **B**.

Therefore, the spectrum estimate for the proposed method is

$$P_1(\theta) = \frac{1}{\sum_{k=1}^{K-1} \left\| \mathbf{a}(\theta)^H \mathbf{B}^{(k)} \right\|^2} \tag{97}$$

where $\mathbf{B}^{(k)}$ is the $k$-th block column of the matrix **B** according to (95).

Clearly, if matrices $\mathbf{B}_1$ and $\mathbf{B}_2$ are defined by (96), the spectral peaks in (97) correspond to the generalized eigenvalues of the matrix pencil $\{\mathbf{C}_i, \mathbf{C}_{i+1}\}$ because

$$P_1(\theta) = \frac{1}{\left\| \mathbf{a}_1(\theta)^H [\mathbf{B}_1 - z(\theta)\mathbf{B}_2] \right\|^2} = \frac{1}{\left\| \mathbf{a}_1(\theta)^H [\mathbf{U} - z(\theta)\mathbf{U}\boldsymbol{\Lambda}] \right\|^2}, \tag{98}$$

where $z(\theta) = \exp(-j2\pi|\Delta|\sin\theta)$; $\mathbf{a}_1(\theta)$ is the steering vector at the aperture of the first subarray.

On the other hand, the estimate in (97) represents the harmonic average $(K-1)$ of the ESPRIT estimates:

$$P_1(\theta) = \frac{1}{\sum_{k=1}^{K-1} \frac{1}{P_{ESP,k}(\theta)}}, \tag{99}$$

where

$$P_{ESP,k}(\theta) = \frac{1}{\left\| \mathbf{a}(\theta)^H \mathbf{B}^{(k)} \right\|^2}. \tag{100}$$

When the array's covariance matrix is block-Toeplitz, estimates (100) from all eigenvector blocks $\mathbf{B}^{(k)}$ coincide. In contrast, non-Toeplitz blocks are unique, meaning harmonic averaging (99) produces an estimate different from standard ESPRIT.

Thus, estimate (97) is a multichannel prediction estimate with minimal entropy (extension of plane wavefronts beyond the antenna aperture) and harmonic averaging of predictions. This makes it a multichannel version of the MUSIC method.

## VII. Unifying Super-Resolution Implementations

Placing super-resolution methods in a common space lets us parameterize spatial spectra. As a result, we can obtain any single-channel prediction (the bottom plane in Fig. 12) through one universal linear filtering process.

From a computational view, we can estimate spatial spectrum parameters in two ways: using the antenna array signal covariance matrix or deriving the parameters directly from the channel signal vectors. Every orthogonalization procedure for sample snapshot vectors directly corresponds to a specific transformation of the covariance matrix. For instance, the Gram-Schmidt method maps to Gaussian elimination, and the modified Gram-Schmidt method maps to Cholesky factorization of a Hermitian matrix. As a baseline, we consider orthogonalization procedures applied to the signal snapshots, while also demonstrating equivalent procedures based on covariance matrices.

### *A. J-orthogonalization*

Spatial spectra can be computed using either covariance matrix operations or direct processing of received signals. The second approach derives spectra through sequential orthogonalization of the antenna array signals, such as Gram-Schmidt orthogonalization, lattice filters, or unitary rotations [15, 29–31]. For clarity, we will demonstrate the J-orthogonalization principle using a modified Gram-Schmidt algorithm [32].

Since the objects to be orthogonalized are vectors of the signal snapshots in the antenna array channels, we redefine the sample covariance matrix for a block of samples. We use the maximum likelihood estimate, omitting the scaling factor as it has no impact on angular coordinate estimation:

$$\mathbf{R} = \mathbf{X}\mathbf{X}^H, \tag{101}$$

where $\mathbf{X} = [\mathbf{X}_1, \mathbf{X}_2, ..., \mathbf{X}_N]^T$ is the matrix of signal samples across the antenna array channels; $\mathbf{X}_n$ is the $K$-dimensional vector representing the complex signal amplitudes at the output of the $n$-th channel, where $K$ is the sample size; and $N$ is the number of channels in the array.

Let **Q** be a non-singular square matrix that linearly transforms the original signals $\mathbf{X}_1, \mathbf{X}_2, ..., \mathbf{X}_N$ into an orthonormal basis $\mathbf{Z}_1, \mathbf{Z}_2, ..., \mathbf{Z}_N$,

$$\mathbf{Z} = \mathbf{Q}\mathbf{X}; \ \mathbf{Z}\mathbf{Z}^H = \mathbf{I}. \tag{102}$$

In this case, the matrix **Q** is the square root of the inverse covariance matrix $\mathbf{R}^{-1}$,

$$\mathbf{Q}\mathbf{X}\mathbf{X}^H\mathbf{Q}^H = \mathbf{I}; \ \mathbf{X}\mathbf{X}^H = (\mathbf{Q}\mathbf{Q}^H)^{-1}; \mathbf{R}^{-1} = \mathbf{Q}\mathbf{Q}^H. \tag{103}$$

Without restrictions on the **Q** matrix, multiple decorrelation transformations satisfy this condition, including the modified Gram-Schmidt orthogonalization algorithm. In this case, the impulse response of the tuned orthogonalizing filter is the lower triangular square root of the inverse covariance matrix,

$$\mathbf{R}^{-1} = \mathbf{L}\mathbf{L}^H, \tag{104}$$

representing a specific version of the matrix **Q**.

The Gram-Schmidt orthogonalization filter yields super-resolution estimates represented by the columns or rows of the lower triangular matrix **L**, including the maximum entropy and Capon estimates.

Next, we introduce the J-orthogonalization procedure, a generalization of the standard orthogonalization. We will use this approach to derive maximum and minimum entropy spatial spectrum estimates.

We call two vectors **x** and **y** J-orthogonal if

$$\mathbf{x}^H\mathbf{J}\mathbf{y} = \mathbf{0}, \tag{105}$$

where **J** is a square matrix.

We supplement the received signal snapshot matrix **X** with a scaled identity matrix $\hat{\sigma}\mathbf{I}$, defining the new matrix as **M**,

$$\mathbf{M} = [\mathbf{X} \quad \hat{\sigma}\mathbf{I}]. \tag{106}$$

Instead of orthogonalizing the rows of **X** as in (102), we enforce J-orthogonality on the rows of the matrix **M**,

$$\mathbf{Z} = \mathbf{Q}\mathbf{M};\ \mathbf{Z}\mathbf{J}\mathbf{Z}^H = \mathbf{I}, \tag{107}$$

where **Q** denotes the transformation matrix, and

$$\mathbf{J} = \begin{bmatrix} \mathbf{I}_K & \mathbf{0} \\ \mathbf{0} & -\mathbf{I}_N \end{bmatrix}, \tag{108}$$

where $\mathbf{I}_K$ and $\mathbf{I}_N$ denote the identity matrices of sizes $K\times K$ and $N\times N$, respectively.

This determines the set of vectors (106), the J-orthogonalization condition (107), and the basis **J** (108). Like standard orthogonalization, various J-orthogonalization procedures exist. While mathematically equivalent, they differ in numerical stability and some other features. We apply the modified Gram-Schmidt algorithm to orthogonalize the rows of **M** in the basis **J** as follows:

$$\mathbf{M}^{(n+1)} = [\mathbf{X}^{(n+1)} \quad \mathbf{H}^{(n+1)}] = \tilde{\mathbf{F}}_n\mathbf{M}^{(n)} = \tilde{\mathbf{F}}_n \ldots \tilde{\mathbf{F}}_1\mathbf{M};\ n = 1, \ldots, N-1 \tag{109}$$

$$\tilde{\mathbf{F}}_n = \left[\begin{pmatrix} 1 & & & & & \mathbf{0} \\ & \ddots & & & & \\ & & 1 & & & \\ & & & (1/\sqrt{\tilde{g}_n}) & & \\ & & & \tilde{f}_{n+1,n} & 1 & \\ & & & \vdots & & \ddots \\ \mathbf{0} & & & \tilde{f}_{N,n} & & 1 \end{pmatrix}\right],$$

$$\tilde{g}_n = \sum_{k=1}^{K}\left|x_{n,k}^{(n)}\right|^2 - \sum_{k=1}^{K}\left|h_{n,k}^{(n)}\right|^2 = \left\|\mathbf{X}_n^{(n)}\right\|^2 - \left\|\mathbf{H}_n^{(n)}\right\|^2, \tag{110}$$

$$\tilde{f}_{i,n} = -\frac{\sum_{k=1}^{K} x_{n,k}^{(n)*}x_{i,k}^{(n)} - \sum_{k=1}^{K} h_{n,k}^{(n)*}h_{i,k}^{(n)}}{\sum_{k=1}^{K}\left|x_{n,k}^{(n)}\right|^2 - \sum_{k=1}^{K}\left|h_{n,k}^{(n)}\right|^2} = \frac{\mathbf{X}_n^{(n)H}\mathbf{X}_i^{(n)} - \mathbf{H}_n^{(n)H}\mathbf{H}_i^{(n)}}{\left\|\mathbf{X}_n^{(n)}\right\|^2 - \left\|\mathbf{H}_n^{(n)}\right\|^2}, \tag{111}$$

$i = n+1, \ldots, N.$

Applying the J-orthogonalization procedure yields J-orthogonal rows for **M** at the filter outputs (Fig. 15):

$$\mathbf{M}^{(N)}\mathbf{J}\mathbf{M}^{(N)H} = \mathbf{I}. \tag{112}$$

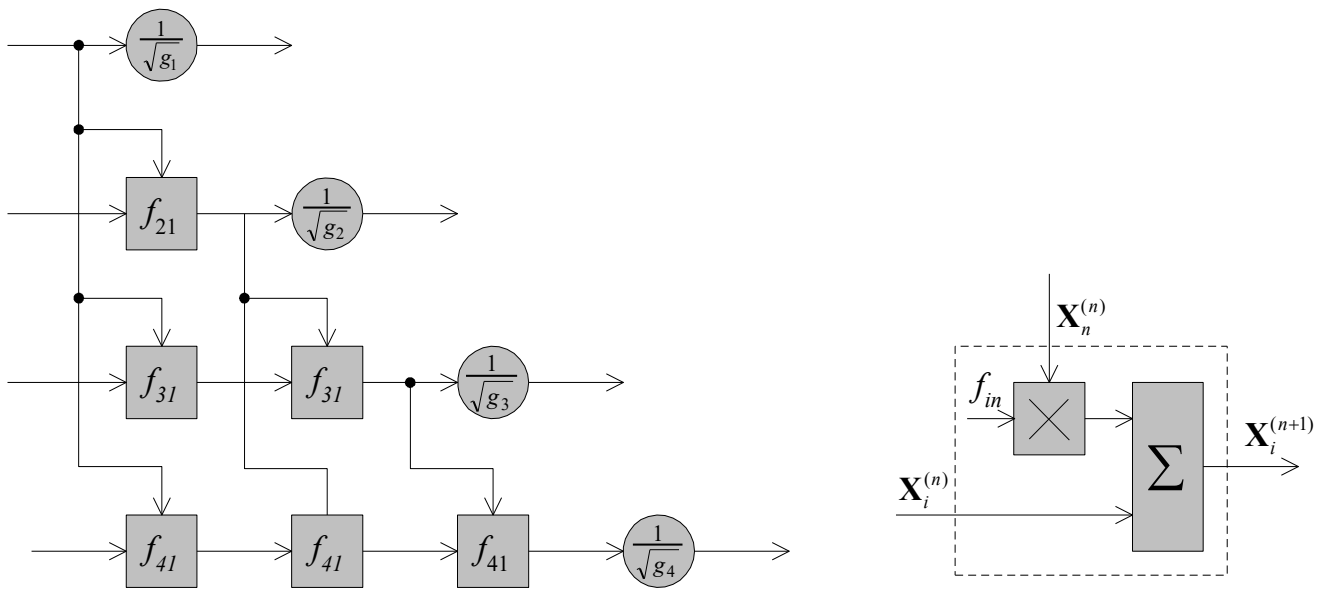

Fig. 15. Block diagram of the Gram-Schmidt J-orthogonalization filter.

The J-orthogonalization filter's impulse response is still represented by a lower triangular matrix:

$$\tilde{\mathbf{L}} = \tilde{\mathbf{D}}\tilde{\mathbf{F}}_{N-1} \ldots \tilde{\mathbf{F}}_2\tilde{\mathbf{F}}_1, \tag{113}$$

where $\tilde{\mathbf{D}} = \mathrm{diag}\left\{\frac{1}{\tilde{g}_1}, \ldots, \frac{1}{\tilde{g}_N}\right\}$.

Suppose, the antiregularization parameter $\hat{\sigma}^2$ differs from all eigenvalues of the sample covariance matrix, ensuring the matrix is non-singular. Then, condition (107) for J-orthogonality can be expressed as:

$$\tilde{\mathbf{L}}\mathbf{M}\mathbf{J}\mathbf{M}^H\tilde{\mathbf{L}}^H = \tilde{\mathbf{L}}(\mathbf{X}\mathbf{X}^H - \sigma^2\mathbf{I})\tilde{\mathbf{L}}^H = \mathbf{I}, \tag{114}$$

Thus, the matrix is not the square root of the inverse covariance matrix $\mathbf{R}^{-1}$, but rather of the matrix $\mathbf{R}_\sigma^{-1}$,

$$\mathbf{R}_\sigma^{-1} = (\mathbf{X}\mathbf{X}^H - \hat{\sigma}^2\mathbf{I})^{-1} = \tilde{\mathbf{L}}^H\tilde{\mathbf{L}}. \tag{115}$$

Here, $\hat{\sigma}^2$acts as an anti-regularization parameter. While regularization adds a value $\hat{\sigma}^2$ to the diagonal ($\mathbf{R}_\sigma = \mathbf{X}\mathbf{X}^H + \hat{\sigma}^2\mathbf{I}$) to improve matrix conditioning, anti-regularization subtracts it ($\mathbf{R}_\sigma = \mathbf{X}\mathbf{X}^H - \hat{\sigma}^2\mathbf{I}$) to do the opposite. The extreme case where $\sigma^2$ equals the noise eigenvalue is also important.

The Gram-Schmidt J-orthogonalization process (111) can also be applied in a regularization mode by changing the subtraction in both the numerator and denominator to addition.

Knowing when J-orthogonalization is feasible is essential. While standard orthogonalization needs the sample covariance matrix **R** being full rank, J-orthogonalization strictly requires that neither **R** nor its leading principal submatrices eigenvalues equal to $\hat{\sigma}^2$. However, setting $\hat{\sigma}^2$ to precisely match a noise eigenvalue of **R** allows J-orthogonalization to generate the eigenvectors needed in projection (minimum-entropy) methods such as Pisarenko and MUSIC.

When the parameter $\hat{\sigma}^2$ is set to the minimum (noise) eigenvalue of the sample covariance matrix $\hat{\sigma}^2 = \lambda_{min}$, then the rank of $\mathbf{R}_\sigma$ corresponds to the number of signal sources, $d$,

$$\mathrm{rank}\{\mathbf{R}_\sigma\} = \mathrm{rank}\{\mathbf{R} - \lambda_{min}\mathbf{I}\} = d. \tag{116}$$

We begin with two lemmas on the normalizing factors under condition (116) before proving the theorem.

**Lemma 1.** For $d = \mathrm{rank}\{\mathbf{R}_\sigma\}$, the factor $g_{d+1}$ vanishes at step ($d$+1) of the J-orthogonalization process (109)–(111),

$$g_{d+1} = \left\|\mathbf{X}_{d+1}^{(d+1)}\right\|^2 - \left\|\mathbf{H}_{d+1}^{(d+1)}\right\|^2 = 0. \tag{117}$$

Proof. Assume for contradiction that $\left\|\mathbf{X}_{d+1}^{(d+1)}\right\|^2 - \left\|\mathbf{H}_{d+1}^{(d+1)}\right\|^2 \neq 0$. This implies that the leading submatrix of

order ($d$+1), given by $\mathbf{R}_{\sigma}^{(d+1)} = \mathbf{R}^{(d+1)} - \hat{\sigma}^2\mathbf{I}_{d+1}$, is of full rank. Consequently, rank$\{\mathbf{R}_\sigma\} \geq d$+1. This directly contradicts the initial assumption that $\mathbf{R}_\sigma = d$, where $d$ is the number of signal sources.

**Lemma 2.** The normalizing factor $g_n$ is nonzero for the first $d$ steps of the J-orthogonalization process,

$$g_n = \left\|\mathbf{X}_n^{(n)}\right\|^2 - \left\|\mathbf{H}_n^{(n)}\right\|^2 \neq 0, n = 1, \ldots, d. \tag{118}$$

Proof. Assume that $\left\|\mathbf{X}_n^{(n)}\right\|^2 - \left\|\mathbf{H}_n^{(n)}\right\|^2 = 0$ for all $n = 1, ..., d$. This implies that the noise-free covariance matrix of the subarray containing the first $n$ antenna elements, $\mathbf{R}_{\sigma}^{(n)} = \mathbf{R}^{(n)} - \hat{\sigma}^2\mathbf{I}_n$, is singular. However, for $n < d$, this matrix is of full rank, regardless of the noise component, which leads to a contradiction.

**Theorem**. Let $\lambda_{\min}$ be the minimum eigenvalue of the matrix $\mathbf{R} = \mathbf{X}\mathbf{X}^H$. The vectors $\mathbf{H}_j^{(d+1)}$ ($j = d$+1, ..., $N$) obtained by the J-orthogonalization (109)–(111) of the rows of the matrix $\mathbf{M} = [\mathbf{X} \quad \hat{\sigma}\mathbf{I}]$ with parameter $\hat{\sigma} = \sqrt{\lambda_{\min}}$ are the eigenvectors of $\mathbf{R}$ corresponding to $\lambda_{\min}$.

Proof. The J-orthogonalization process (109)–(111) acts as a linear filtering procedure for the vectors in packet $\mathbf{M}$. At the filter input, the square block $\mathbf{H}^1$ of matrix $\mathbf{M}$ is a scalar matrix $\mathbf{H}^1 = \sigma\mathbf{I}$. After the $n$-th step, it becomes a transformation matrix representing the filter's impulse response—up to the coefficient $\hat{\sigma}$. Consequently, vectors $\mathbf{X}_j^{(d+1)}$ equal the product of the input signal packet $\mathbf{X}$ and the corresponding impulse response,

$$\mathbf{X}_j^{(d+1)T} = \frac{1}{\hat{\sigma}}\mathbf{H}_j^{(d+1)T}\mathbf{X}. \tag{119}$$

Because the normalizing factor $g_{d+1}$ (117) equals zero, it follows

$$\mathbf{X}_j^{(d+1)H}\mathbf{X}_j^{(d+1)} = \mathbf{H}_j^{(d+1)H}\mathbf{H}_j^{(d+1)}. \tag{120}$$

Using (119) in (120), we find

$$\frac{1}{\sigma}\mathbf{H}_j^{(d+1)H}\mathbf{X}\mathbf{X}^H\mathbf{H}_j^{(d+1)} = \mathbf{H}_j^{(d+1)H}\mathbf{H}_j^{(d+1)}, \mathbf{H}_j^{(d+1)H}\mathbf{R}\mathbf{H}_j^{(d+1)} = \sigma^2\left\|\mathbf{H}_j^{(d+1)}\right\|^2, \tag{121}$$

$$\frac{\mathbf{H}_j^{(d+1)H}\mathbf{R}\mathbf{H}_j^{(d+1)}}{\left\|\mathbf{H}_j^{(d+1)}\right\|^2} = \hat{\sigma}^2. \tag{122}$$

The left side of the last equation is the Rayleigh ratio. Its minimum value equals the smallest eigenvalue of the numerator's Hermitian matrix,

$$\min_{\mathbf{H}_j^{(d+1)}}\left\{\frac{\mathbf{H}_j^{(d+1)H}\mathbf{R}\mathbf{H}_j^{(d+1)}}{\left\|\mathbf{H}_j^{(d+1)}\right\|^2}\right\} = \lambda_{min}, \tag{123}$$

As $\hat{\sigma}^2$ in (122) is the minimum eigenvalue of $\mathbf{R}$ by assumption, it follows that $\mathbf{H}_j^{(d+1)}$ is the associated eigenvector, completing the proof.

Thus, the vectors $\mathbf{H}_j^{(d+1)}$ ($j = d$+1, ..., $N$) represent the noise eigenvectors of the covariance matrix, assuming the parameter $\hat{\sigma}$ is the square root of the minimum noise eigenvalue. This enables the implementation of the Pisarenko and MUSIC minimum entropy methods within the J-orthogonalization process.

*B. Bridging Minimum and Maximum Entropy Spatial Spectra*

We show that a J-orthogonalization filter yields the angular super-resolution estimates for all single-channel prediction methods on the lower bound of the estimate space cube (Fig. 12). These methods include Pisarenko, MUSIC, Capon, and maximum entropy. First, we rewrite each spatial spectrum expression to fit an adaptive linear filter. We omit normalizing factors since they do not change the DOA estimates, unless doing so causes confusion.

The maximum entropy spatial spectrum estimate is defined as:

$$P_{ME}(\theta) = \frac{1}{\left|\mathrm{a}(\theta)^H \boldsymbol{l}_{*1}\right|^2}, \tag{124}$$

where $\mathbf{a}(\theta)$ is the steering vector, and $\boldsymbol{l}_{*1}$ is the first column of the lower triangular matrix $\mathbf{L}$ from the decomposition $\mathbf{R}^{-1} = \mathbf{L}\mathbf{L}^H$.

The Capon estimate is given by:

$$P_{CP}(\theta) = \frac{1}{\sum_{k=1}^{N}\left|\mathrm{a}(\theta)^H \boldsymbol{l}_{*k}\right|^2}, \tag{125}$$

where $\boldsymbol{l}_{*k}$ is the $k$-th column of the matrix $\mathbf{L}$

The Pisarenko spatial spectrum estimate is:

$$P_{PSR}(\theta) = \frac{1}{\left|\mathrm{a}(\theta)^H \mathbf{e}_1\right|^2}, \tag{126}$$

where $\mathbf{e}_1$ is the unique noise eigenvector of the covariance matrix.

The MUSIC estimate is:

$$P_{MUSIC}(\theta) = \frac{1}{\sum_{k=1}^{N-d}\left|\mathrm{a}(\theta)^H \boldsymbol{e}_k\right|^2}, \tag{127}$$

where $\boldsymbol{e}_k$ is the $k$-th noise eigenvector of the covariance matrix and $d$ is the number of signal sources.

Passing steering vectors through a tuned Gram-Schmidt orthogonalization filter yields maximum entropy or Capon spatial spectrum estimates. These estimates come from inverting the output power of the $N$-th filter output (124) for maximum entropy, or the sum of all filter outputs (125) for Capon.

Another method directly obtains the desired $\boldsymbol{l}_{*k}$ vectors in the J-orthogonalization filter with parameter $\hat{\sigma} = 1$, adjusting the filter coefficients according to the standard Gram-Schmidt orthogonalization algorithm. The 'tails' of the packet $\mathbf{M}$ are excluded from the coefficient calculations; instead, they are filtered along with the packet's training portion and then substituted into the maximum entropy (124) and Capon (125) spectrum estimates correspondingly:

$$P_{ME}(\theta) = \frac{1}{\left|\mathrm{a}(\theta)^H \mathbf{H}_N^{(N)}\right|^2}, \tag{128}$$

$$P_{CP}(\theta) = \frac{1}{\sum_{k=1}^{N}\left|\mathrm{a}(\theta)^H \mathbf{H}_k^{(N)}\right|^2}. \tag{129}$$

Next, we consider obtaining minimum entropy estimates based on the eigenvectors of the noise subspace during J-orthogonalization. To transition from maximum entropy and Capon estimates to the minimum entropy estimates of the Pisarenko and MUSIC, it is sufficient to set the antiregularization parameter $\hat{\sigma}^2$ in the J-orthogonalization filter (109)–(111) equal to the minimum (noise) eigenvalue $\lambda_{\min}$ of the covariance matrix.

The precise equality of the antiregularization parameter to the noise eigenvalue is less critical than it seems. Choosing an underestimated value only shifts the spatial spectrum toward higher entropy. As shown below, the spectrum changes smoothly across the antiregularization parameter from $\lambda_{\mathrm{m}}$ to zero, moving from the minimum to the maximum entropy spectrum. The following estimate of the spatial spectrum using the outputs of the J-orthogonalization filter corresponds to the Pisarenko method:

$$P_{PSR}(\theta) = \frac{1}{\left|\mathbf{a}(\theta)^H \mathbf{H}_N^{(d+1)}\right|^2}. \tag{130}$$

The MUSIC method uses $N-d$ noise subspace vectors from the covariance matrix. We assume the noise eigenvalues are equal, meaning the minimum eigenvalue has a multiplicity of $N-d$. These noise vectors appear as $N-d$ vectors $\mathbf{H}_j^{(d+1)}$ ($j = d$+1, ..., $N$) at the output of the $d$-th filter cascade (Fig. 15) using J-orthogonalization. Because they correspond to a multiple eigenvalue, these vectors are not mutually orthogonal yet. However, they fully define the noise subspace. We orthogonalize them across the remaining $N-d-1$ filter cascades using the standard Gram-Schmidt algorithm without changing the filter structure. This yields mutually orthogonal vectors defining the MUSIC spatial spectrum:

$$P_{MUSIC}(\theta) = \frac{1}{\sum_{k=1}^{N-d}\left|\mathbf{a}(\theta)^H \mathbf{H}_{(d+j)\perp}^{(N)}\right|^2}. \tag{131}$$

By applying the J-orthogonalization technique, we derive four primary super-resolution estimates situated at the opposite extremes of the entropy scale. We will next develop intermediate-entropy estimates that bridge the gap between the Pisarenko and maximum entropy spectrums, as well as and the MUSIC and Capon spectrums.

Producing spectrum estimates based on intermediate entropy values in the J-orthogonalization filter is equivalent to obtaining Capon estimates. However, the antiregularization parameter $\hat{\sigma}^2$ must range between zero and the noise eigenvalue of the covariance matrix matrix $0 < \hat{\sigma}^2 < \lambda_{\min}$.

Varying the number of averaged predictions gives us every super-resolution estimate in the single-channel prediction plane (Fig. 12):

$$P(\theta)|_{\sigma,K} = \frac{1}{\sum_{k=1}^{K}\left|\mathbf{a}(\theta)^H \mathbf{H}_{N-K+k}^{(N)}\right|^2}, \tag{132}$$

where $K$ denotes the number of averaged predictions.

We have demonstrated how to obtain all super-resolution estimates, from minimum to maximum entropy, including the optional harmonic averaging of predictions within the J-orthogonalization filter. Fig. 16 illustrates the derivation of spectrum estimates for the four super-resolution methods on the single-channel prediction plane (Fig. 12), including any intermediate estimates. In all scenarios, the filter inputs consist of a composite packet **M**, which contains signal samples from the antenna array channels **X**, and a scalar square matrix σ**I**.

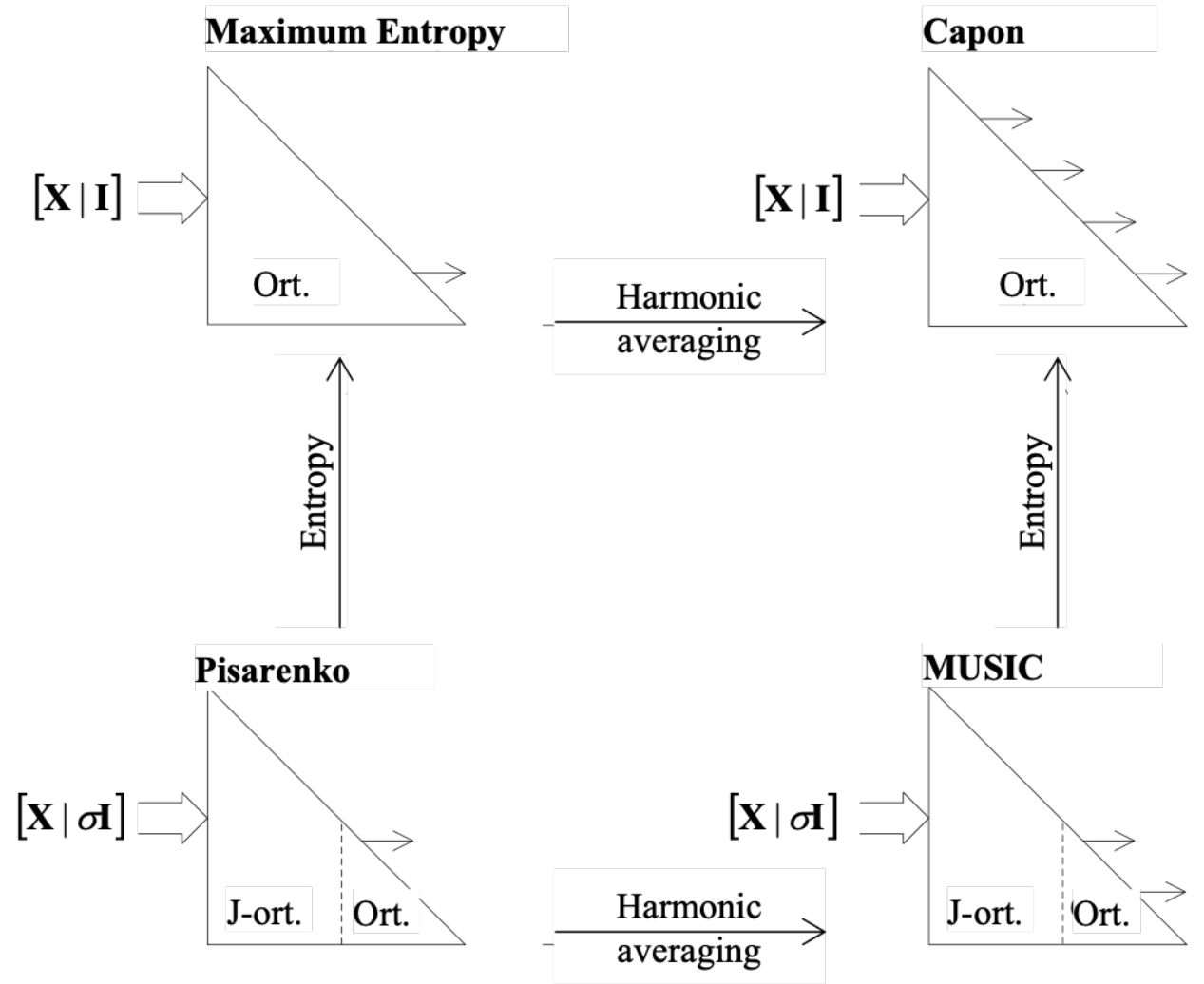

Fig. 16. Mapping of super-resolution methods to J-orthogonalization processes.

When the parameter $\sigma$ equals one and the filter executes a standard Gram-Schmidt orthogonalization on the receiving signals **X**, filtering the entire augmented matrix **M** = [**X** | **I**], we obtain maximum entropy and Capon estimates.

Pisarenko and MUSIC estimates are derived by setting the filter's antiregularization parameter equal to the covariance matrix's minimum eigenvalue. As filtering proceeds, the last $N-d$ rows of the submatrix $\hat{\sigma}\mathbf{I}$ become eigenvectors of the noise subspace.

Spatial spectrum estimates can be computed directly from the prediction parameter vectors of the J-orthogonalization filter. Alternatively, they can be obtained by passing the steering vectors $\mathbf{a}(\theta_i)$ through the tuned filter, an approach illustrated in Fig. 17.

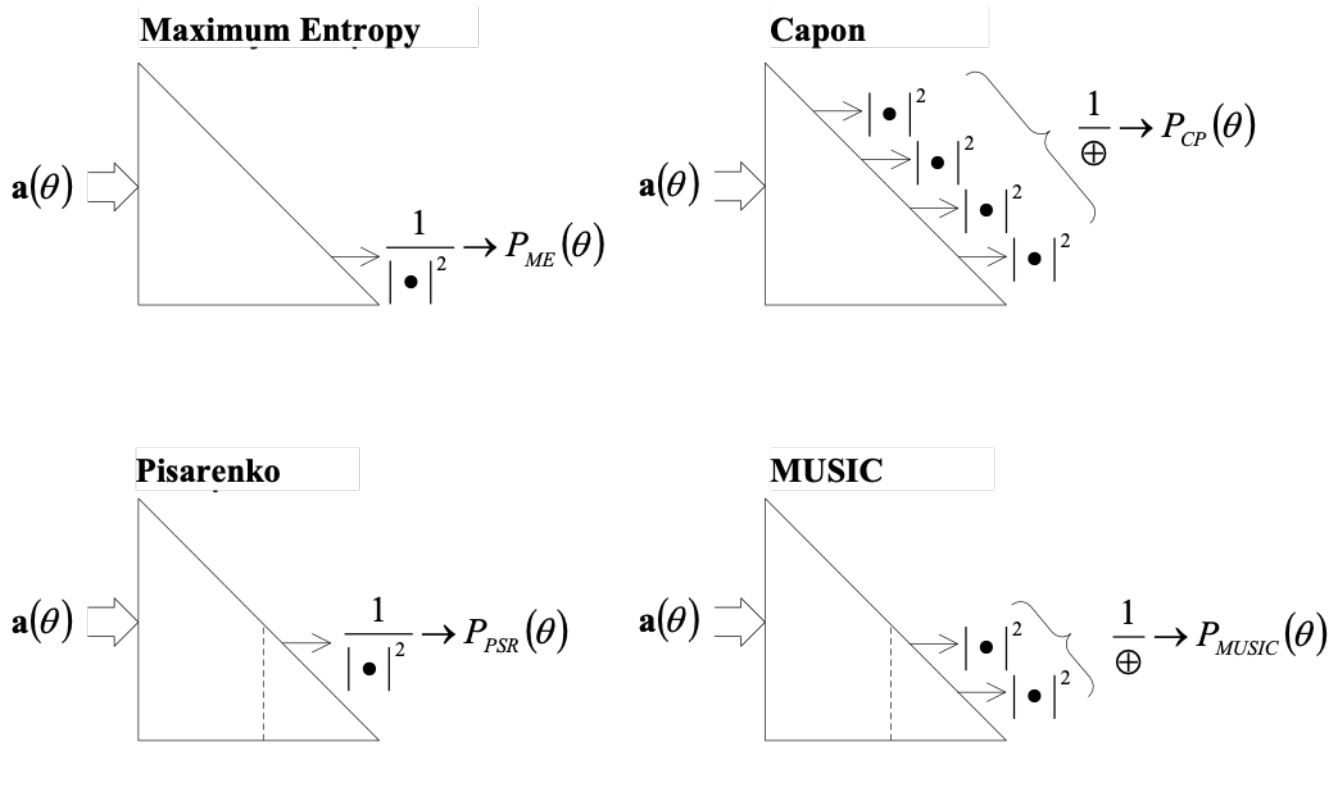

Fig. 17. Deriving spatial spectra with a tuned J-orthogonalization filter.

We demonstrate how a J-orthogonalization filter can yield maximum entropy, Pisarenko, Capon, and MUSIC super-resolution estimates, and their intermediate variants. Our simulation features a 16-element ULA receiving signals from four uncorrelated Gaussian sources in additive uncorrelated Gaussian noise. The angular coordinates of these sources are set to -0.2, -0.16, 0.1, and 0.25, with corresponding powers of 10, 5, 10, and 1 dB.

Fig. 18 illustrates the spatial spectra generated by the J-orthogonalization filter for the maximum entropy, Capon, Pisarenko, and MUSIC estimators.

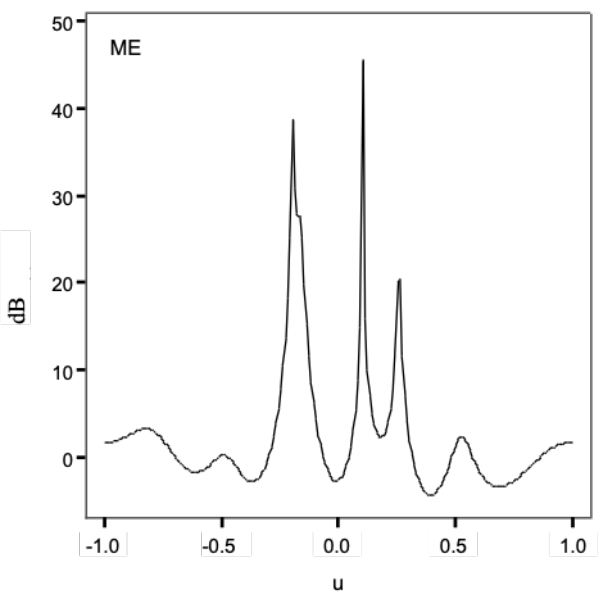

(a) maximum entropy

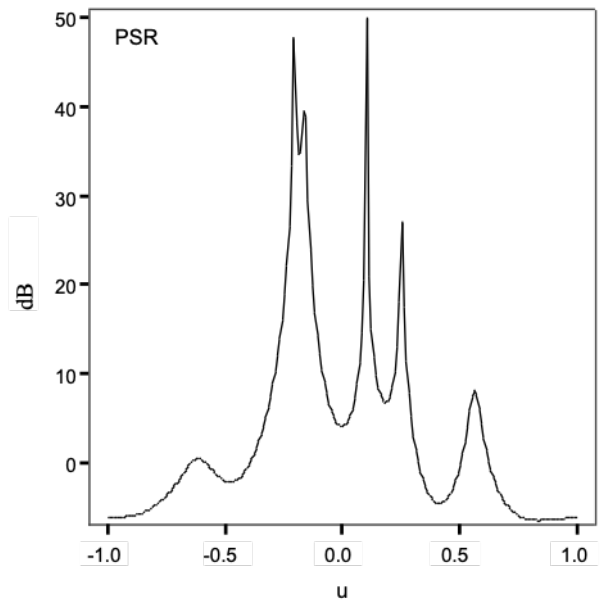

(b) Pisarenko

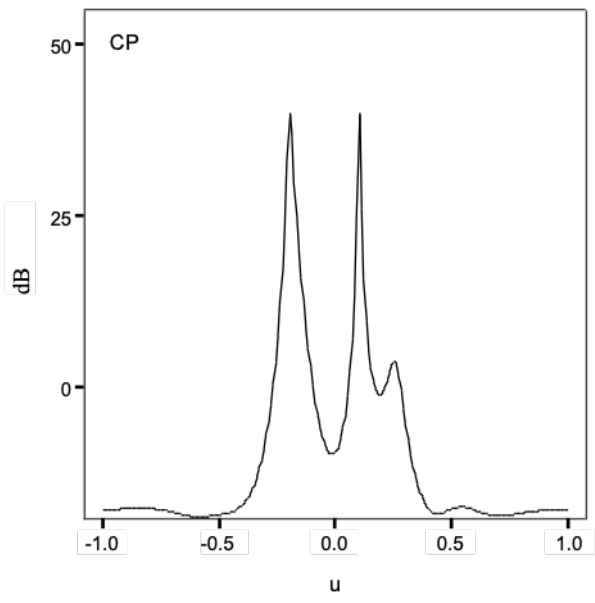

(c) Capon

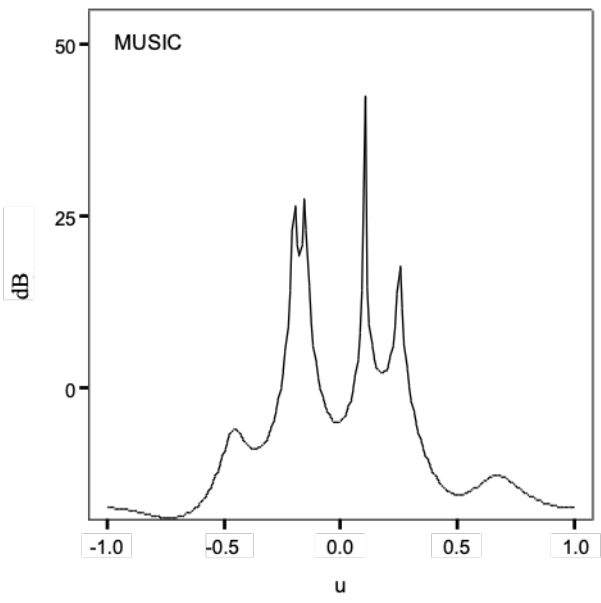

(d) MUSIC

Fig. 18. Spatial spectra estimated in the J-orthogonalization filter.

Figures 19 and 20 display the spatial spectrum estimates across a range of intermediate entropy values from maximum to minimum, without and with harmonic averaging, respectively. The entropy axis shows how the spatial spectrum changes across different values of the antiregularization parameter ($\sigma^2$ = 0, 0.1, 0.3, 0.5, 0.7, 0.9, 1.0). Position 7 in the foreground represents maximum entropy ($\sigma^2 = 0$), while position 1 in the background represents minimum entropy ($\sigma^2 = 1$). While the true covariance matrix has a minimum multiple noise eigenvalue of exactly 1, the sample covariance matrix yields varying noise eigenvalues ranging from 0.8 to 1.13. This means the Pisarenko and MUSIC methods use approximate extensions of the minimum-entropy spatial correlation function rather than exact extensions—which are equivalent to noise subspace projectors—their spatial spectra may differ slightly from Fig. 19.

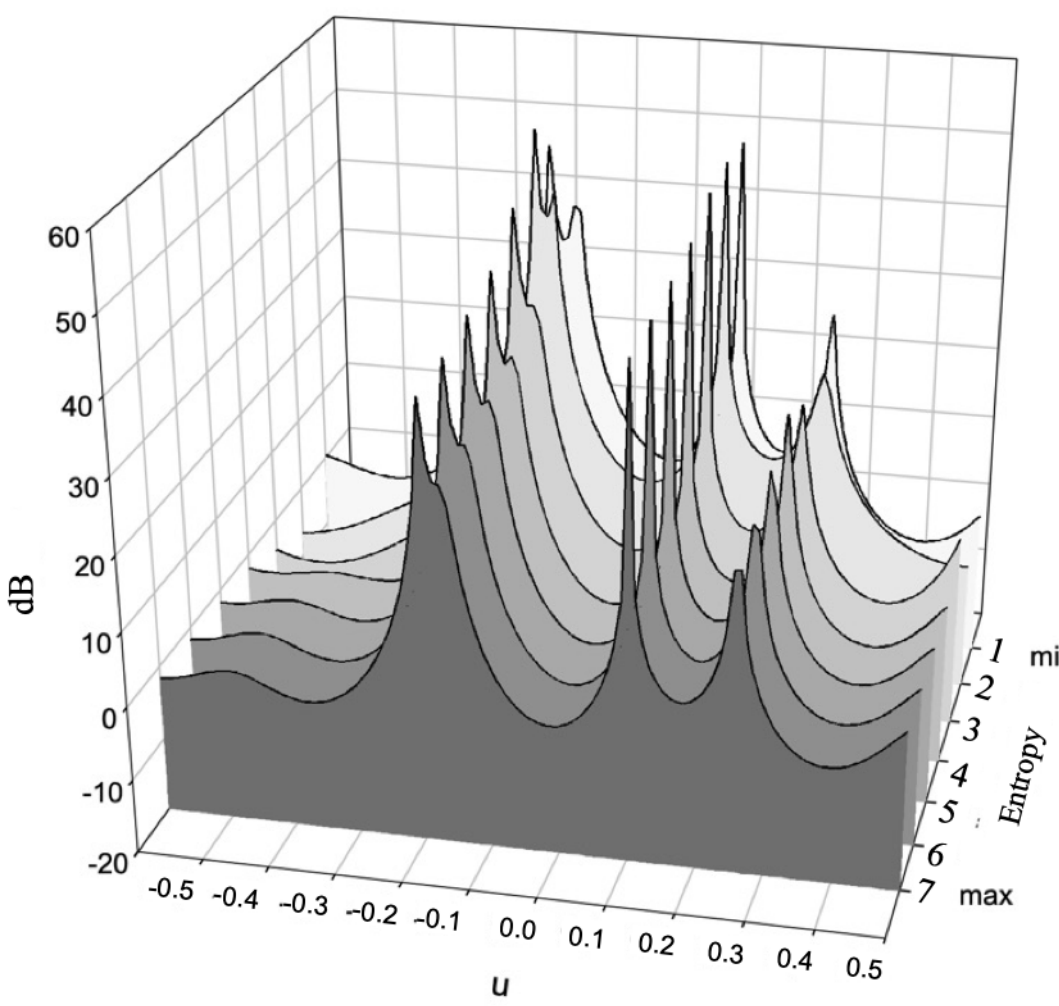

Fig. 19. Spatial spectra ranging from minimum to maximum entropy (Pisarenko to maximum entropy) estimated using the J-orthogonalization filter.

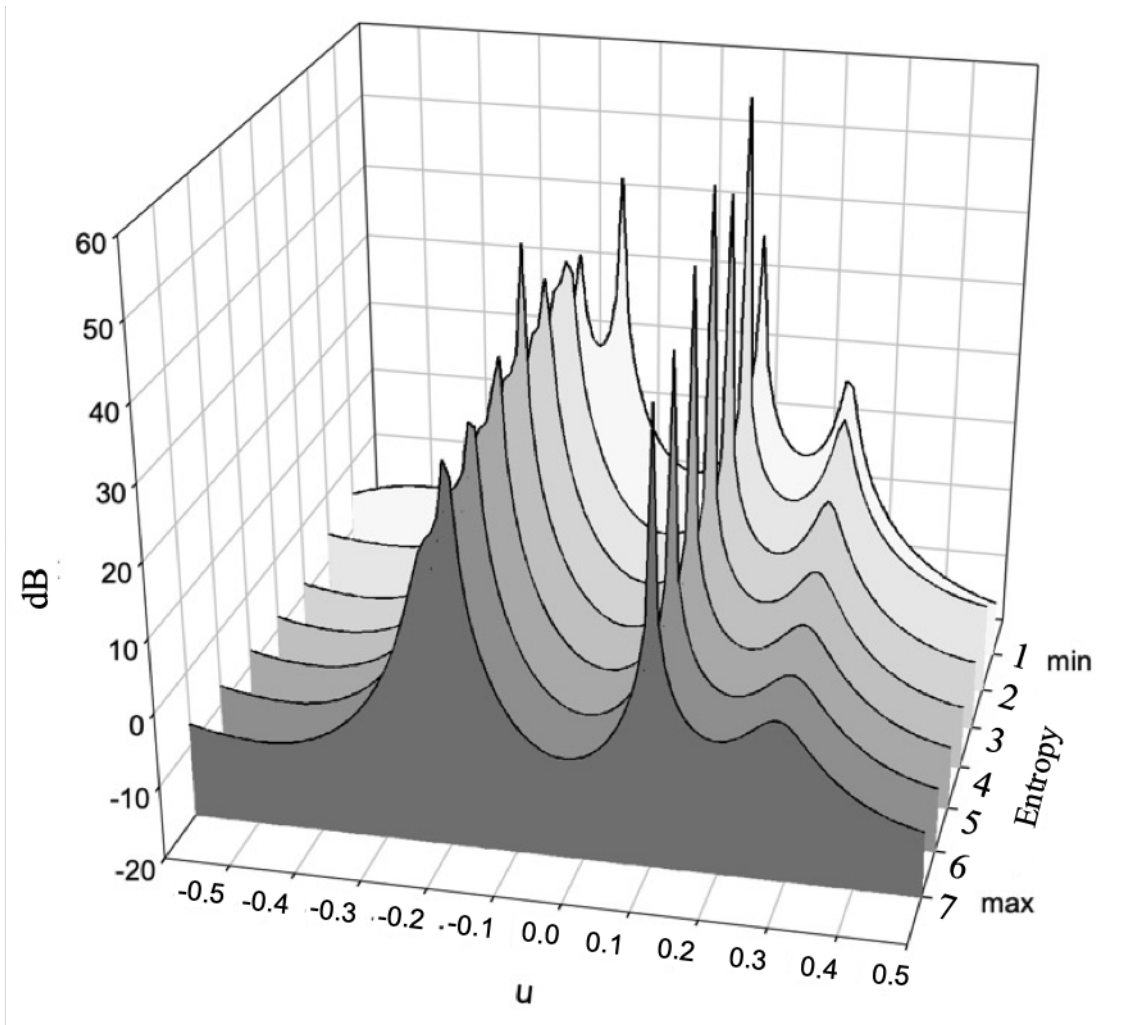

Fig. 20. Spatial spectra ranging from minimum to maximum entropy with harmonic averaging (MUSIC to Capon) estimated using the J-orthogonalization filter.

Fig. 19 shows that the angular resolution increases as the antiregularization parameter approaches the noise eigenvalue (i.e., as entropy decreases toward Pisarenko estimate). This improvement is evident from the distinct resolution of the closely spaced sources at coordinates –0.2 and –0.16.

Fig. 20 shows the spatial spectra obtained from the J-orthogonalization filter using harmonic averaging of the estimates. The Capon spectrum appears in the foreground, characterized by maximum entropy (zero antiregularization parameter, $\sigma^2 = 0$). Moving along the entropy axis, the final plot converges toward the MUSIC minimum entropy estimate (the antiregularization parameter $\sigma^2 = 1$).

## VIII. Conclusions

Super-resolution methods can be organized within a coordinate space defined by three core factors: entropy, estimate averaging, and multichannel subarray prediction. This framework not only structures existing methods but also accommodates any intermediate estimates. Building on this, our approach leverages a universal J-orthogonalization algorithm to simultaneously generate spectral estimates across multiple super-resolution methods with virtually no extra computational overhead. Consequently, it enables the adaptive selection or combination of super-resolution methods to optimize performance under varying signal-to-noise ratios and array calibration conditions.